\documentclass[12pt]{article}
\usepackage{natbib, color, amssymb, amsmath, graphicx, amsthm, bm, multirow, subcaption, mathrsfs, fancyhdr, adjustbox, scalerel, lscape, soul, mathtools, cite, xr, verbatim, enumerate}
\usepackage[labelsep=period,labelfont=bf]{caption}
\usepackage[ruled]{algorithm2e}

\def\T{^T}
\def\var{\textnormal{Var}}
\def\cov{\textnormal{Cov}}
\newcommand{\bbf}{\mathbf{b}}
\def \onebf {\bm{1}}

\newcommand{\Y}{\mathbf{Y}}
\newcommand{\X}{\mathbf{X}}
\newcommand{\Z}{\mathbf{Z}}
\newcommand{\betabf}{\bm\beta}
\newcommand{\Gammabf}{\bm\Gamma}
\newcommand{\thetabf}{\bm\theta}
\newcommand{\varepsilonbf}{\bm\varepsilon}
\newcommand{\dif}{\mathrm{d}}

\newcommand{\alphabf}{\bm\alpha}
\newcommand{\Omegabf}{\bm\Omega}
\newcommand{\inv}{^{-1}}
\newcommand{\Sigmabf}{\bm\Sigma}
\newcommand{\detz}{\mathrm{
	det_0
}}
\newcommand{\I}{\mathbf{I}}
\def \zerobf {\bm{0}}
\newcommand{\spann}{\mathrm{span}}
\newcommand{\mh}{{-\frac{1}{2}}}

\newcommand{\E}{\mathbb{E}}
\newcommand{\tr}{\mathrm{tr}}
\newcommand{\V}{\mathbf{V}}

\newcommand{\vecc}{\text{vec}}

\newcommand{\Pro}{\mathbf{P}}
\newcommand{\Q}{\mathbf{Q}}
\newcommand{\subtildeSigmabfepsilonbf}{\scaleto{\widetilde\Sigmabf_{\scaleto{\varepsilon}{3pt}}}{6pt}}
\newcommand{\subSigmabfepsilonbf}{\scaleto{\Sigmabf_{\scaleto{\varepsilon}{3pt}}}{6pt}}
\theoremstyle{definition}
\newtheorem{condition}{Condition}

\newtheorem{lemma}{Lemma}

\newtheorem*{corollary}{Corollary}
\newtheorem{proposition}{Proposition}
\newtheoremstyle{named}{}{}{\itshape}{}{\bfseries}{}{.5em}{\thmnote{#3's }#1}
\theoremstyle{named}
\makeatletter
\newcommand*\bigcdot{\mathpalette\bigcdot@{.5}}
\newcommand*\bigcdot@[2]{\mathbin{\vcenter{\hbox{\scalebox{#2}{$\m@th#1\bullet$}}}}}
\makeatother

\makeatletter
\newcommand*{\indep}{%
	\mathbin{%
		\mathpalette{\@indep}{}%
	}%
}
\newcommand*{\@indep}[2]{%
	\sbox0{$#1\perp\m@th$}
	\sbox2{$#1=$}
	\sbox4{$#1\vcenter{}$}
	\rlap{\copy0}
	\dimen@=\dimexpr\ht2-\ht4-.2pt\relax
	\kern\dimen@
	{#2}%
	\kern\dimen@
	\copy0 
} 
\makeatother

\def\var{\textnormal{Var}}

\def \zerobf {\bm{0}}
\def \onebf {\bm{1}}

\title{Mixed Effects Envelope Models\protect\footnote{Yuyang Shi and Linquan Ma contributed equally to this work.}}
\date{}
\author{Yuyang Shi$^{1*}$, Linquan Ma$^{2*}$ Lan Liu$^{3}$\\
	$^{1}${\small School of Industrial and Systems Engineering, Georgia Institute of Technology}\\
	$^{2}${\small Department of Statistics, University of Wisconsin - Madison}\\
	$^{3}${\small School of Statistics, University of Minnesota at Twin Cities}}

\begin{document}
\maketitle

\begin{abstract}
	When multiple measures are collected repeatedly over time, redundancy typically exists among responses.   The envelope method was recently proposed to reduce the dimension of responses without loss of information in  regression with multivariate responses. It can gain substantial efficiency over the standard least squares estimator. In this paper, we generalize the envelope method to mixed effects models for longitudinal data with possibly unbalanced design and time-varying predictors. We show that our model provides more efficient estimators than the standard estimators in mixed effects models. Improved accuracy and efficiency of the proposed method over the standard mixed effects model estimator are observed in both the simulations and the Action to Control Cardiovascular Risk in Diabetes (ACCORD) study.

	\textbf{Keywords:} Envelope method, mixed effects model, sufficient dimension reduction, efficiency gain.
\end{abstract}

\section{Introduction}
\subsection{Literature review}
Over the past three decades, an increasing amount of literature has emerged on the topic of sufficient dimension reduction (SDR). \citet{li1991sliced} proposed the sliced inverse regression to reduce the dimension of the predictors. That is, assuming the response only depends on a linear combination of the predictors, one regresses the predictors $\X$ against the response $\Y$ to circumvent any model-fitting process. \citet{cook1998regression} defined the central subspace as the subspace with the minimal dimension
such that the response is independent of the predictors given the projection of the predictors onto the space. Other SDR methods include but not limited to sliced average variance estimation \citep{cook1991comment},  principal Hessian direction \citep{li1992principal}, contour regression \citep{li2005contour}, inverse regression estimation  \citep{cook2005sufficient}, directional regression \citep{li2007directional}, likelihood-acquired directions \citep{cook2009likelihood}, discretization-expectation estimation \citep{zhu2010sufficient}, non-elliptically distributed predictors \citep{li2009dimension,dong2010dimension},  dimension reduction based on canonical correlation \citep{fung2002dimension,zhou2008dimension}, and average partial mean estimation \citep{zhu2010dimension}. 
However, the aforementioned methods all focus on the dimension reduction of predictors with univariate response. 

Recently, \citet{cook2010envelope} proposed a new sufficient dimension reduction method called the envelope method to reduce the dimension of {\it responses} in multivariate regression. 
Specifically, \citet{cook2010envelope} considered the following multivariate linear regression model 
\begin{equation}\label{eq: regular_multivariate_lm}
\Y_i = \bm\alpha + \bm\beta \X_i + \bm\varepsilon_i,
\end{equation}
where $i$ indicates the $i^{th}$ individual, $\Y_i,~ \bm\varepsilon_i\in \mathbb{R}^r$, $\X_i\in\mathbb{R}^p$ and the parameter of interest is $\bm\beta$.

The key idea of the  envelope method is to assume the existence of redundancy in responses that do not contribute to the estimation of $\betabf$ in model \eqref{eq: regular_multivariate_lm}, so that the estimation of $\betabf$ is more efficient by leveraging this condition.  
\citet{cook2010envelope} assumes that there exists an orthogonal matrix $({\bm\Gamma},{\bm\Gamma}_0)\in \mathbb{R}^{r\times r}$,  where ${\bm\Gamma}\in \mathbb{R}^{r\times u}$ and  ${\bm\Gamma}_0\in \mathbb{R}^{r\times (r-u)}$, with $0\leq u\leq r$ satisfying the following conditions:

\begin{condition}
	$\bm\Gamma_0^T\Y_i \indep \mathbf X_i$;
\end{condition}
\begin{condition}
	${\bm\Gamma}^T \Y_i \indep {\bm\Gamma}_0^T \Y_i \mid \X_i$,
\end{condition}
\noindent where $\indep$ indicates independence.
Under Conditions 1 and 2, $\Gammabf_0^T\Y_i$ is redundant for a fixed effect regression \citep{cook2010envelope}. 
The $\Sigmabf_{\varepsilon}$-envelope is uniquely defined to be the smallest subspace satisfying these conditions. Once the basis $\widehat{\bm \Gamma}$ is obtained, the envelope estimator is obtained by projecting the ordinary least square estimator onto the estimated envelope space. 
\citet{cook2010envelope} showed that the envelope estimator can achieve efficiency gain over the OLS estimator. Following the definition in their paper, we define the variance of $\bm\Gamma^T_0 \Y_i\mid \X_i$ as the material part variance, and $\bm\Gamma^T_0 \Y_i$ as the immaterial part variance. The efficiency gain will be substantial if the variation of the immaterial part is relatively large as compared with that of the material part. 

The envelope methods have been developed in different settings, including response envelope \citep{cook2010envelope}, partial envelope \citep{su2011partial}, inner envelope \citep{su2012inner}, scaled envelope \citep{cook2013scaled}, predictor envelope \citep{cook2013envelopes}, reduced rank envelope \citep{cook2015envelopes}, simultaneous envelope \citep{cook2015simultaneous}, model-free envelope \citep{cook2015foundations}, 
and tensor envelope \citep{li2017parsimonious}. 

Longitudinal data, also known as panel data, collects repeated measurements of the same subjects over time.  As a distinctive feature of longitudinal data, measures that are collected repeatedly over time  are typically correlated and redundancy typically exists among responses. Hence, reducing data to lower dimensions can improve efficiency while still preserving all relevant information on regression.  Additionally, data may be unbalanced in the sense that subjects are not measured at the same time points. Moreover, the predictors may depend on time and may have different trajectories over time across individuals. For example, in a study of the effect of smoking on body weight, the number of cigarettes smoked per day may stay the same for some individuals but not for others. These features distinguish the longitudinal data from the cross-sectional data in terms of appropriate analyses. However, the study of sufficient dimension reduction for longitudinal data is quite limited. \citet{pfeiffer2012sufficient} developed first-moment sufficient dimension reduction techniques to replace the original predictors with longitudinal nature. \citet{bi2015sufficient} applied the quadratic inference function to longitudinal data sufficient dimension reduction. The literature is even more scarce on dimension reduction in mixed effects models with high-dimensional response. Some early work has been done by \citet{zhou2010reduced} using a reduced rank model for spatially correlated hierarchical functional data and by \citet{hughes2013dimension} using sparse reparameterization to reduce spatial confounding.

In this paper, we propose a mixed effects envelope model. Similar to the standard mixed effects model, the variability of each observation are composed of the within-individual variability and the between-individual variability. The mixed effects envelope model recovers the distribution of the unobserved between-individual random coefficient and reduces noises in the within-individual variation. The mixed effects envelope model inherits both the efficiency gain of the standard envelope model and the flexibility of the standard mixed effects model. Specifically, our methods result in more efficient estimators than those from the standard mixed effects models and can be used for unbalanced data as well as with  time-varying predictors.

\subsection{Notation}

Consider a study with $n$ individuals with each individual being measured at a total of $J_i$ time points, where $i = 1, \ldots, n$ and $j=1,\ldots,J_i$. Let $J_{\bigcdot} = \sum_{i=1}^{n}J_i$ denote the total number of the observations across time. For an individual $i$ at time $j$, let $\mathbf Y_{ij}\in \mathbb{R}^r$ denote the responses of length $r$, let $\mathbf X_{ij}\in \mathbb{R}^p$ denote the vector of predictors of length $p$, and let $\mathbf Z_{ij}\in \mathbb R^q$ denote the vector of predictors of length $q$. Predictors $\mathbf X_{ij}$ and $\mathbf Z_{ij}$  can either be stochastic or nonstochastic. Let $\mathbf Y_i=(\mathbf Y_{i1},\ldots \mathbf Y_{iJ_i})$ denote the responses, and $\mathbf X_i=(\mathbf X_{i1},\ldots \mathbf X_{iJ_i})$, $\mathbf Z_i=(\mathbf Z_{i1},\ldots \mathbf Z_{iJ_i})$ denote the predictors for one individual at all time points. Let $\mathbb R^{r\times p}$ denote the class of all matrices with size $r\times p$. Let $\mathbb S^{r\times r}$ denote the class of all symmetric positive definite matrices of size $r$. Let $\mathbf I_r$ denote the identity matrix of size $r$. Let vec($\cdot$) denote the vectorization of a matrix by stacking the columns of the matrix on top of one another, and let vech($\cdot$) denote the vectorization of the unique part of each column that lies on or below the diagonal. Let $\dagger$ denote the Moore-Penrose inverse. Also, let $\mathbf E_r \in \mathbb{R}^{r^2\times r(r+1)/2}$ to be the expansion matrix such that $\mathrm{vec}(\cdot) = \mathbf E_r \mathrm{vech}(\cdot)$ and $\mathbf C_r \in \mathbb{R}^{r(r+1)/2\times r^2}$ to be the contraction matrix such that $\mathrm{vech}(\cdot) = \mathbf C_r \mathrm{vec}(\cdot)$. Let $\widetilde \Y_i=\mathrm{vec}(\Y_i)$ denote the vectorized responses. Let $\mathbf A\otimes\mathbf B$ denote the kronecker product between $\mathbf A$ and $\mathbf B$. All the population covariance matrices in this paper are positive definite. We use $\spann(\mathbf A)$ to denote the span of the column vectors of $\mathbf A$.

\subsection{Organization of the paper}
We organize this paper as follows.  In Section \ref{sec: prelim}, we give the standard mixed effects model and discuss a special case where classic envelope can be directly applied. In Section \ref{sec: inference}, we propose the mixed effects envelope model as well as provide a graphical illustration of our method. We further illustrate our proposed method in the simulations in Section \ref{sec: simulation} and data analysis in Section \ref{sec: data_analysis}.  We conclude with a brief discussion in Section \ref{sec: discussion}. 

\section{Preliminary}\label{sec: prelim}
\subsection{Mixed effects model}

Consider the mixed effects model
\[\mathbf Y_{ij} = \bm \alpha + \bm\beta \mathbf X_{ij} + \mathbf b_i\mathbf Z_{ij} + \bm\varepsilon_{ij},\]
where $\bm\alpha\in\mathbb{R}^r$ is the intercept, $\bm\beta\in \mathbb{R}^{r\times p}$ denotes the coefficient for the fixed effects and $\mathbf b_i\in \mathbb{R}^{r\times q}$ denotes the random coefficients.  
Assume vec($\mathbf b_i$) identically and independently follows $N(\bm{0}, {\bm\Sigma}_{\mathbf b})$, where $\bm\Sigma_{\mathbf b} \in \mathbb S^{qr\times qr}$. The residual error $\bm\varepsilon_{ij}$ identically and independently follows $N(\mathbf{0},{\bm\Sigma}_{\bm\varepsilon})$ for $i =1,\ldots,n$ and $j = 1,\ldots,J_i$, where ${\bm\Sigma}_{\bm\varepsilon}\in \mathbb S^{r\times r}$. The normality of random effect and error are assumed here for simplicity. We extend our result when they have finite $(4+\delta)$-th moment in Section \ref{sec: inference}. The random effect $\mathbf b_i$ is assumed to be independent from the residual error $\bm\varepsilon_{ij}$, i.e., $\mathbf b_i\indep \bm\varepsilon_{ij}$. The variance due to random coefficient $\bm\Sigma_{\mathbf b}$ is the between-subject variability and the variance due to the error $\bm\Sigma_{\bm\varepsilon}$ is the within-subject variability.  Let $\mathbf A_{ij}= \mathbf Z_{ij}^T\otimes\mathbf I_r$, then $\mathbf b_i\mathbf Z_{ij} = (\mathbf Z_{ij}^T\otimes\mathbf I_r )\mathrm{vec}(\mathbf b_i)=\mathbf A_{ij}\mathrm{vec(\mathbf b_i)}$. The covariance of responses across time for the same individual is  correlated $\cov(\mathbf Y_{ij}, \mathbf Y_{ij'}\mid\mathbf X_{ij},\mathbf X_{ij'},\mathbf Z_{ij},\mathbf Z_{ij'}) = \mathbf A_{ij}{\bm\Sigma}_{\mathbf b}\mathbf A_{ij'}^T$ if $j\neq j'$, and $\var(\mathbf Y_{ij}\mid\mathbf X_{ij},\mathbf Z_{ij}) = \mathbf A_{ij}{\bm\Sigma}_{\mathbf b}\mathbf A_{ij}^T + \bm\Sigma_{\bm\varepsilon}$.  
Let $\bm\varepsilon_i=(\bm\varepsilon_{i1},\ldots \bm\varepsilon_{iJ_i})$, we can rewrite the model above in a matrix form as
\begin{equation}\label{eq: mixed_model}
\mathbf Y_{i} = \bm \alpha\otimes \onebf^{T}_{J_i}+ \bm\beta \mathbf X_{i} + \mathbf b_i\mathbf Z_{i} + \bm\varepsilon_{i}.
\end{equation}


\noindent Model \eqref{eq: regular_multivariate_lm} is a special case of model \eqref{eq: mixed_model}  when $\bm\Sigma_{\bbf}=\zerobf$, and $J_i = 1$ for $i = 1,\ldots, n$.


\subsection{Classic envelope model for a special case of longitudinal data}
The classic envelope method can be applied to longitudinal data in a special case: when the data is balanced, the predictors do not vary with time, and random slopes are not included in the model (random intercepts are included). 
We will show that under this setting, the mixed effects model naturally contains an envelope structure over the observations across time.
Under this setting, $J_i = J$ for any $i$. Also, if we assume $\mathbf X_i$ does not vary with time, then \eqref{eq: mixed_model} can be written as
\begin{equation}\label{eq: classic_noenv_long_bal}
\widetilde\Y_i = \onebf_{J}\otimes\alphabf + (\mathbf 1_{J}\otimes\betabf) \X_i + \widetilde \varepsilonbf_i,
\end{equation} 

\noindent where $\widetilde\varepsilonbf_i\in\mathbb{R}^{rJ}$ i.i.d follows $N(\zerobf,\widetilde\Sigmabf_{\varepsilon})$ and $\widetilde\Sigmabf_{\varepsilon} = \mathbf I_J\otimes \bm\Sigma_{\mathbf b} + \Sigmabf_{\bm\varepsilon}$. This model is a standard multivariate model, hence we can impose an envelope model on it.
Let $\mathcal{E}_{\subtildeSigmabfepsilonbf}(\widetilde{\mathcal{B}})$ denote the $\widetilde\Sigmabf_{\varepsilon}$-envelope for $\widetilde {\mathcal B}$, where $\widetilde {\mathcal B} = \text{span}(\mathbf 1_{J}\otimes\betabf)$. The structure of $\mathcal{E}_{\subtildeSigmabfepsilonbf}(\widetilde{\mathcal{B}})$ 
is given in the following proposition and corollary.
\begin{proposition}\label{prop: classic_env_long_bal_same_beta}
	Under model \eqref{eq: classic_noenv_long_bal}, the basis for $\mathcal{E}_{\subtildeSigmabfepsilonbf}(\widetilde{\mathcal{B}})$ is $\mathbf 1_J \otimes \bm\Phi$, where $\bm\Phi$ is the basis for $\mathcal{E}_{(\bm\Sigma_{\bm\varepsilon}/J + \bm\Sigma_{\mathbf b})}(\mathcal{B})$.
	
\end{proposition}
\begin{corollary}
	Under model \eqref{eq: classic_noenv_long_bal}, the dimension of $\mathcal{E}_{\subtildeSigmabfepsilonbf}(\widetilde{\mathcal{B}})\subseteq \mathbb R^{rJ}$ cannot exceed $r$.
\end{corollary}
Intuitively, 
although the repeated measures from the same individual are correlated, 
because neither the fixed effects nor the random effects change over time, we can reduce the dimension of responses by averaging each individual over different time points. That is, model \eqref{eq: mixed_model} 
naturally results in combinations of the responses of dimension $r(J-1)$ that do not contribute to the regression. This results in a $\widetilde\Sigmabf_{\varepsilon}$-envelope $\mathcal{E}_{\subtildeSigmabfepsilonbf}(\widetilde{\mathcal{B}})$ with envelope dimension no greater than $r$ rather than $rJ$.  

Proposition \ref{prop: classic_env_long_bal_same_beta} presents a simple but important observation:
if the true model is a mixed effects model but  instead we fit a standard multivariate linear regression, even we have a reduced dimension from $rJ$ to $r$ by the envelope method, we do not gain additional efficiency. This is because the failure to leverage the mixed effects model structure creates redundancy.  
Such an observation naturally leads us to explore an envelope model that can incorporate the mixed effects model structure to gain further efficiency.

\section{The mixed effects envelope model}\label{sec: inference}

\subsection{Conditions}\label{subsec: mixed_model}
Now, we propose the mixed effects envelope model. The key requirement of the classic envelope method is the existence of some linear combination of the responses that do not contribute to the regression. With longitudinal data, because both $\X_i$ and $\Z_i$ are observed predictors, it may seem natural to extend Conditions 1 and 2 by replacing $\X_i$ with $(\X_i,\Z_i)$ as 

\vspace{1.5mm}
\noindent\textit{Condition }{{1$^{\circ}$.}}\label{assump: 1circ}
$\bm\Gamma_0^T\Y_i \indep (\X_i,\Z_i)$;

\vspace{1.5mm}
\noindent\textit{Condition }{{2$^{\circ}$.}}\label{assump: 2circ}
${\bm\Gamma}^T \Y_i\indep {\bm\Gamma}_0^T \Y_i\mid\X_i,\Z_i$. 
\vspace{1.5mm}

It has been shown that the standard envelope Conditions 1 and 2 are equivalent to the reparameterization $\mathrm{span}(\bm \beta)\subseteq \mathrm{span}(\bm \Gamma)$ and ${\bm\Sigma}_{\bm\varepsilon} = {\bm\Gamma}{\bm\Omega}{\bm\Gamma}^T+{\bm\Gamma}_0{\bm\Omega}_0{\bm\Gamma}_0^T$, where  $\Omegabf=\bm\Gamma^T\bm\Sigma_{\bm\varepsilon}\bm\Gamma$, and $\Omegabf_0=\bm\Gamma_0^T\bm\Sigma_{\bm\varepsilon}\bm\Gamma_0$. Unlike Conditions  1 and 2 which impose conditions on population parameters,  Conditions  1$^{\circ}$ and 2$^{\circ}$ are equivalent to requiring certain relationships between $\Z_i$ and parameters as shown in the proposition below. In general, Conditions  1$^{\circ}$ and 2$^{\circ}$ are hard to satisfy because their validity is contingent on the observed value of $\Z_i$ in the sample. 

\begin{proposition}\label{prop: conditional_distribution}
	Conditions 1$^{\circ}$ and 2$^{\circ}$ hold under model \eqref{eq: mixed_model} if and only if 
	$\Gammabf_0\T\betabf=\zerobf$, $\Z_i\otimes\bm\Gamma_0 = \zerobf$, and $\mathbf \I_{J_i}\otimes(\bm\Gamma\T\bm\Sigma_{\bm\varepsilon}\bm\Gamma_0)+(\Z_i\T\otimes\bm\Gamma\T)\bm\Sigma_{\mathbf b}(\Z_i\otimes\bm\Gamma_0)=\zerobf$.
\end{proposition}


To modify Condition  1$^{\circ}$, we want to find a condition that reduces to Condition 1 when there is no random effect. Recall under \eqref{eq: regular_multivariate_lm}, $\vecc(\Gammabf_0\T\Y_i)$ only depends on predictors through $\betabf\X_i$ in the mean and to have its distribution free of $\betabf$ is the same as to have $\bm\Gamma_0^T\Y_i\indep\bm\X_i$. In other words, Condition 1 can be equivalently expressed as $\E({\bm\Gamma}_0\T \Y_i\mid\X_i)=\E( {\bm\Gamma}_0\T \Y_i)$ under linear model. However, under model \eqref{eq: mixed_model}, $\vecc(\Gammabf_0\T\Y_i)$ depends on the predictors $(\X_i,\Z_i)$ through both mean and variance. Notice that the parameter of interest $\betabf$ only involves in the mean of $\vecc(\Gammabf_0\T\Y_i)$.  Thus, this motivates us to relax the distributional independence between $\bm\Gamma_0\T\Y_i$ and $(\X_i,\Z_i)$ to be just mean independence, i.e., $\E({\bm\Gamma}_0\T \Y_i\mid\X_i, \Z_i)=\E( {\bm\Gamma}_0\T \Y_i)$ so that this condition reduces to Condition 1 when there is no random effect.  

To modify Condition  2$^{\circ}$, we also want to find a condition that reduces to Condition 2 in the absence of the random effect. Note 
$\vecc(\Gammabf\T\Y_i)\mid\vecc(\Gammabf_0\T\Y),\X_i,\Z_i,\bbf_i\sim N(\bm\mu^{***}, \bm\Sigma^{***}),$ 
where
\begin{align*}
	\bm\mu^{***} = \vecc(\bm\Gamma\T\bm\beta\X_i+\bm\Gamma\T\mathbf b_i\mathbf Z_i)+\I_{J_i}\otimes\left\{\bm\Gamma\T\bm\Sigma_{\bm\varepsilon}\bm\Gamma_0(\bm\Gamma_0\T\bm\Sigma_{\bm\varepsilon}\bm\Gamma_0)\inv \right\}\left\{\vecc(\Gammabf_0\T\Y_i) -\vecc(\Gammabf_0\T\bm\beta\X_i+\bm\Gamma_0\T\mathbf b_i\mathbf Z_i)\right\},
\end{align*}  and $\bm\Sigma^{***} = \I_{J_i}\otimes(\Gammabf\T\bm\Sigma_{\bm\varepsilon}\inv\Gammabf)\inv.$ 
That is, if we conditional on both predictors and random effects, the conditional independence between $\Gammabf\T\Y_i$ and $\Gammabf_0\T\Y_i$ is equivalent to $\bm\Gamma\T\bm\Sigma_{\bm\varepsilon}\bm\Gamma_0=0$, i.e., $\Gammabf$ reduces $\Sigmabf_{\bm \varepsilon}$. This condition reduces to Condition 2 when there is no random effect. 

Thus, to develop the mixed effects envelope model, we assume 

\vspace{1mm}
\textit{Condition }{{1$^*$.}}\label{assump: 1*}
$\E({\bm\Gamma}_0\T \Y_i\mid\X_i, \Z_i)=\E( {\bm\Gamma}_0\T \Y_i)$,

\vspace{1mm}
\textit{Condition }{{2$^*$.}}\label{assump: 2*}
${\bm\Gamma}\T \Y_i\indep {\bm\Gamma}_0\T \Y_i\mid\X_i,\Z_i,\bbf_i$. 
\vspace{1mm}


As mentioned, in the absence of random effects, Condition 1$^{*}$ and 2$^{*}$ will reduce to Condition 1 and 2 under the linear model.  Conditions 1$^{*}$ and 2$^{*}$ can be viewed as  extensions of Conditions 1 and 2 for longitudinal data. However, unlike the classic envelope condition \citep{cook2010envelope}, Condition 1$^{*}$ only requires the expectation of ${\bm\Gamma}_0\T \Y_i\mid\X_i, \Z_i$ and $ {\bm\Gamma}_0\T \Y_i$ to be the same. 
The motivation of Condition 1$^{*}$ is that instead of imposing the redundancy of ${\bm\Gamma}_0\T \Y_i$ on its entire distribution, we just assume the redundancy on its mean, which is easier to satisfy. 

Condition 2$^{*}$ assumes the independence between ${\bm\Gamma}\T \Y_i$ and ${\bm\Gamma}_0\T \Y_i$ conditional on predictors $(\X_i,\Z_i)$, as well as on the unobservable $\bbf_i$. Equivalently, the redundancy of responses is within individuals rather than across. In other words, Condition 2$^{*}$ excludes the possibility of ${\bm\Gamma}_0\T \Y_i$ contributing to the regression through a correlation with ${\bm\Gamma}\T \Y_i$ given any individual, although such correlation may be present in the population.  Here, different from the original envelope model, under model \eqref{eq: mixed_model}, $ \Sigmabf_{\bm\varepsilon}$ is the variance of outcomes given predictors and random effects, which is only the within-subject variation not including the between-subject variation. Thus, when the study is balanced, even when the classic envelope model does not have much efficiency gain, the mixed effects envelope may achieve substantial efficiency gain: 
decomposing part of the variability into material and immaterial variability may be possible even when  decomposing the total variability may not be possible. The idea of using part of parameters to form an envelope was also adopted in the development of partial envelope \citep{su2011partial}, where part of parameters in the mean model are used. 

Other than having clear interpretations,  Conditions 1$^{*}$ and 2$^{*}$ also facilitate the reparameterization of the original parameters in  \eqref{eq: mixed_model}. Under Condition 1$^{*}$, $\E({\bm\Gamma}_0\T \Y_i\mid\X_i, \Z_i)$ is free of $\betabf$, which indicates $\mathrm{span}(\bm \beta)\subseteq \mathrm{span}(\bm \Gamma)$. Additionally, since Condition 2$^{*}$ conditions on the random effects, we have  ${\bm\Sigma}_{\bm\varepsilon} = {\bm\Gamma}{\bm\Omega}{\bm\Gamma}\T+{\bm\Gamma}_0{\bm\Omega}_0{\bm\Gamma}_0\T$. 
We define the smallest reducing subspace that satisfies Conditions 1$^*$ and 2$^*$ as the mixed effects envelope, or $\Sigmabf_{\varepsilon}$-mean envelope and write it as $\bar{\mathcal{E}}_{\subSigmabfepsilonbf}({\mathcal{B}})$.
Under Conditions 1$^*$ and 2$^*$, model (\ref{eq: mixed_model}) can be written as: 
\begin{equation}
\label{eq: mix_env}
\begin{aligned}
\mathbf Y_{ij} &= \bm\alpha + \bm\Gamma\bm \eta  \mathbf X_{ij} + \mathbf b_i\mathbf Z_{ij} +  \bm\varepsilon_{ij},
\end{aligned}
\end{equation} 
where $\bm\varepsilon_{ij}$ identically and independently follows $N(\mathbf{0},{\bm\Sigma}_{\bm\varepsilon})$, and ${\bm\Sigma}_{\bm\varepsilon} = {\bm\Gamma}{\bm\Omega}{\bm\Gamma}\T+{\bm\Gamma}_0{\bm\Omega}_0{\bm\Gamma}_0\T$. 

Under the mixed effects envelope model \eqref{eq: mix_env}, the number of variational independent parameters changes from $r + rp + r(r+1)/2 + qr(qr+1)/2$ to $r + up + r(r+1)/2 + qr(qr+1)/2$. Since $u\leq r$, the number of parameters of mixed effects envelope model is no more than the standard mixed effects model. When $J$ is large, the number of parameters in the mixed effects envelope model \eqref{eq: mix_env} can be substantially fewer than those in the standard envelope model, but there is no general relationship between the number of parameters in these two envelope models. 

Under Conditions 1$^{*}$ and 2$^{*}$ and model \eqref{eq: mixed_model}, the covariance $\var(\widetilde \Y_i\mid\X_i,\Z_i)$ has a specific heteroscedastic error structure $\var(\widetilde \Y_{i}\mid\X_i,\Z_i) = \mathbf A_{i}{\bm\Sigma}_{\mathbf b}\mathbf A_{i}\T+\I_{J_i}\otimes\bm\Sigma_{\bm\varepsilon}$, where $\mathbf A_{i}=\mathbf Z_{i}^T\otimes \I_r$. Another heteroscedastic error model was considered in \citet{su2013estimation}, where the predictors are only indicators for the subpopulation and individuals in the same subpopulation have the same distribution. Following \citet{su2013estimation}, \citet{park2017groupwise} generalized the multivariate envelope mean model to groupwise envelope regression models with heteroscedastic error. In their setting, the envelope is assumed to be the intersection of subspaces that contains columns of all the coefficients across populations. Under model \eqref{eq: mixed_model}, it is possible that $\Z_i$ is different for all individuals, then we have $n$ single individual subpopulations. This situation cannot be directly handled in their framework. 

\subsection{Graphical illustration}\label{sec: graph}
Before diving into the estimation details of the mixed effects envelope model, we 
first provide a graphical illustration of the classic envelope, the standard mixed effects  estimator and our mixed effects envelope estimator, under the mixed effects model.
We generate the outcomes $\Y_i$ from  (1) with $\Z_{ij}=1$. To compare the classic envelope with the mixed effects envelope, we only consider the setting
where data is balanced and the predictors are time-invariant.

Consider two groups of individuals with $X_i=0$ and $1$ respectively. 
We generate $n = 2000$ individuals with $J_i = 5$ observation for each group. For individual $i$ at time point $j$, we generate a bivariate response $\Y_{ij}=(Y_{ij1},Y_{ij2})\T$ from the mixed effects model (1) with  $\bm\alpha = \bm 0$ and $\bm \beta = (-7.07, 7.07)\T$. 
We are interested in examining the mean group difference. 

\begin{figure}[!ht]
	
	\caption{Graphical illustration of the  standard mixed effects estimator and the mixed effects envelope estimator, with 
		a random intercept. Individuals of the two groups are represented by cross dots ($X=0$) and triangles ($X=1$). The scatter points in Figure \ref{demon_plot_mixed2}a demonstrate the original data, whereas Figure \ref{demon_plot_mixed2}b and \ref{demon_plot_mixed2}c demonstrate the data of $\Y-\bbf$ as if $\mathbf b$ is observed. Solid curves are from the EM-type estimates, and the dashed curves are from the estimates when $\mathbf b$ is given.}
	\centering
	\includegraphics[width=.7\textheight]{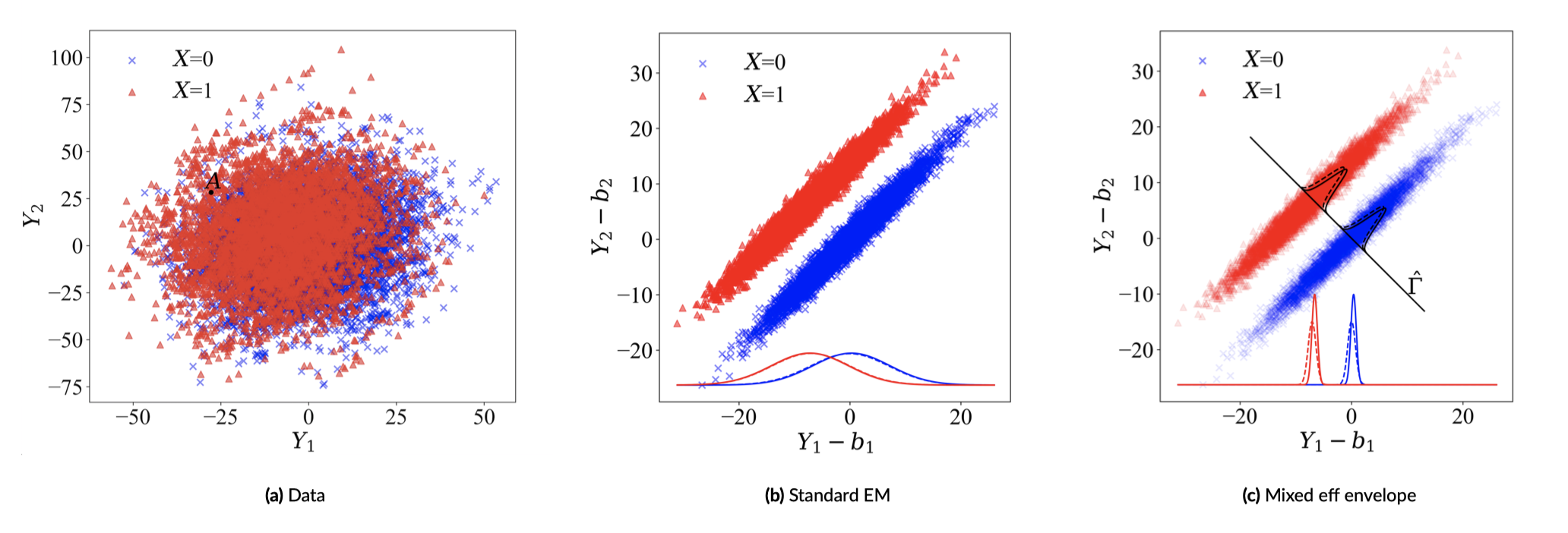}
	\label{demon_plot_mixed2}
\end{figure}

Figure \ref{demon_plot_mixed2}a presents the raw data, where we directly implement the OLS method and the classic envelope model. 
The OLS estimator is $(-7.54,6.38,-7.17,6.83,$ $-7.23,6.66,-7.81,6.28,\\-7.68,6.27)^T$ 
with the MSE 1.10. Model \eqref{eq: classic_noenv_long_bal} ignores the fact that some responses are repeated measures over time and does not distinguish them from different measures collected at the same time point. As a result, the OLS estimator of model \eqref{eq: classic_noenv_long_bal} is 
relatively inefficient. 

By applying the classic envelope method, the estimated envelope dimension is $\widehat{u} = 2$. The envelope estimate for the group difference is $(-7.46, 6.52, -7.49, 6.49, -7.41, 6.53, -7.50, 6.49,$ $-7.57, 6.39)^T$ 
with the MSE 0.93. 
The classic envelope removes some redundancy in the responses as compared with the OLS estimator from model \eqref{eq: classic_noenv_long_bal}. However, as we show below, such redundancy can easily be removed by incorporating the mixed effects model structure. 

Figure \ref{demon_plot_mixed2}b shows the performance of the estimator  under model (1) using the expectation-maximization (EM) algorithm, a standard mixed effects model estimator. For comparison, we keep the OLS estimate of $\Y_{ij} - \bbf_{i}$ in two groups  when $\mathbf b$ is assumed observed (dashed curve in Figure \ref{demon_plot_mixed2}b) as a benchmark. The solid curves denote the estimated distribution of $\Y_{ij} - \bbf_{i}$ using the standard EM. The density curves related to the EM algorithm are not simply projections of scattered points obtained by subtracting $\bbf$, and the scatter points just provide intuitions.  The EM algorithm separates the between-subject variability from the within-subject variability when calculating the point estimates, hence, the solid and dashed density curves almost overlap completely. The group difference using the standard EM is $(-7.48, 6.48)^T$ with 
the MSE 0.95. The MSE is similar to that of the classic envelope in (5), which confirms that incorporating the mixed effects model already eliminates some noise in the repeated measures. 

Figure \ref{demon_plot_mixed2}c illustrates the performance of our mixed effects envelope method. 
The solid curves at the bottom are obtained by applying our method on the data set in Figure \ref{demon_plot_mixed2}a. 
Unlike the standard EM algorithm which only recovers the between-subject variability, our method additionally reduces the within-subject variability. The estimated envelope dimension $\widehat u = 1$. The group difference estimated from our method is $(-6.97, 6.99)^T$ 
with the MSE 0.40. The solid and dashed curves are similar, indicating that our method provides a similar point estimate as the standard envelope estimator with random effects subtracted. While the classic envelope estimator has almost no efficiency gain over the standard EM estimator (similar MSEs), our mixed effects envelope method achieves substantial efficiency gain over the standard EM estimator (the MSE ratio is only about 0.4) and even more efficiency gain over the OLS estimator from model (4) (the MSE ratio is about 0.36). This shows that by leveraging both the mixed effects model and the envelope structure, our method may achieve a much greater amount of efficiency gain as compared with either method. 

\subsection{Maximum likelihood estimation} \label{MLE}
Under the reparameterization implied by Conditions 1$^*$ and 2$^*$, we first investigate the likelihood function given the observed data. Recall that $\mathbf A_{i}=\mathbf Z_{i}^T\otimes\I_r$. We have 
\begin{equation} \label{obs_data_likelihood}
\begin{aligned}
L({\bm \theta}, \mathbf Y; \mathbf X, \mathbf Z)&\propto\prod_{i = 1}^n
|\mathbf I_{J_i}\otimes\bm\Sigma_{\bm\varepsilon} + \mathbf A_i\bm\Sigma_{\mathbf b}\mathbf A_i\T|^{-1/2}\exp\bigg\{-\dfrac{1}{2}\{\mathrm{vec}(\mathbf Y_i) - \bm\alpha\otimes\mathbf {1}_{J_i} - \mathrm{vec}(\bm\beta\X_i)\}\T\\
&\quad(\mathbf I_{J_i}\otimes\bm\Sigma_{\bm\varepsilon} + \mathbf A_i\bm\Sigma_{\mathbf b}\mathbf A_i\T)\inv\{\mathrm{vec}(\mathbf Y_i) - \bm\alpha\otimes\mathbf {1}_{J_i} - \mathrm{vec}(\bm\beta\X_i)\}\bigg\},
\end{aligned}
\end{equation}

\noindent where ${\bm \theta}=(\bm\alpha, {\bm\Gamma},\bm\eta,{\bm\Omega},{\bm\Omega}_0,{\bm\Sigma}_{\mathbf b})$, $\bm\beta={\bm\Gamma}\bm\eta$, ${\bm\Sigma}_{\bm\varepsilon} ={\bm\Gamma}{\bm\Omega}{\bm\Gamma}\T+{\bm\Gamma}_0{\bm\Omega}_0{\bm\Gamma}_0\T$ and $\propto$ denotes proportional to. The formula above is obtained by square completion and the Woodbury matrix identity. 
As the likelihood function $L({\bm \theta},\mathbf Y; \mathbf X, \mathbf Z)$ have a complicated form in ${\bm \theta}$, the MLE  of model \eqref{eq: mix_env} does not have a closed form in general.

In order to obtain the maximum likelihood estimate, we combine the EM-algorithm and the envelope structure. The resulting algorithm is not trivial since the parameters are not element-wise identifiable and random effects are not observable. Due to space constraints, we only briefly describe the steps here and relegate the technical details of the algorithm 
in the Supplementary Materials. For any predetermined envelope dimension $u$, we start with an initial value for all parameters. We calculate the E-step and then, during the M-step, we decompose the expectations from the E-step such that all the other parameters can be optimized individually given $\bm\Gamma$, and then we optimize over $\bm\Gamma$. We iterate such EM process till convergence. 

We adapt the BIC in the classic envelope models \citep{eck2017weighted} to estimate $u$ in the mixed effects envelope model. Under model \eqref{eq: mix_env}, BIC is $-2l(\widehat{\bm \theta};\mathbf Y\mid\mathbf X, \mathbf Z) + \log(J_{\bigcdot})pu$,
where $l(\widehat{\bm \theta};\mathbf Y\mid\mathbf X, \mathbf Z)$ is the $\log$ of 
the likelihood $L({\bm \theta}, \mathbf Y; \mathbf X, \mathbf Z)$ given in \eqref{obs_data_likelihood}. The penalty coefficient in BIC is $\log(J_{\bigcdot})$ rather than $\log(n)$ to take the longitudinal feature of data into consideration \citep{jones2011bayesian}. Also, the likelihood in BIC is the observed data likelihood rather than the full data one. We summarize the mixed effects envelope algorithm in the appendix.

\subsection{Efficiency Gain}
We discuss the asymptotic variance of the mixed effects envelope estimator. The parameters of the envelope model is vector $\bm\phi = (\bm\eta, \bm\Gamma, \bm\Omega, {\bm\Omega}_0, \bm\Sigma_{\mathbf b})$. A more rigorous notation is $\bm\phi = (\mathrm{vec}(\bm\eta), \mathrm{vec}(\bm\Gamma),$ $ \mathrm{vech}(\bm\Omega), \mathrm{vech}({\bm\Omega}_0), \mathrm{vech}(\bm\Sigma_{\mathbf b}))$. We omit the vectorization notations here. We are interested in the property of the parameter $\bm\beta$, ${\bm\Sigma}_{\bm\varepsilon}$ and $\bm\Sigma_{\mathbf b}$, which can be viewed as functions of $\bm\phi$. Generally, we have $\mathbf h(\bm \phi) = (\bm \beta, {\bm\Sigma}_{\bm\varepsilon}, \bm\Sigma_{\mathbf b}) = ({\bm\Gamma} \bm\eta, {\bm\Gamma} {\bm\Omega}{\bm\Gamma}\T + {\bm\Gamma}_0{\bm\Omega}_0{\bm\Gamma}_0\T, \bm\Sigma_{\mathbf b}) = (h_1(\bm \phi), h_2(\bm \phi), h_3(\bm \phi))$. Let ${\bm \theta}=\mathbf h(\bm\phi)$, $\widehat{{\bm \theta}}_{mix\cdot env}$ and $\widehat{{\bm \theta}}_{mix\cdot em}$ denote the mixed effects envelope and standard EM estimates under \eqref{eq: mixed_model}. The asymptotic variance of our estimator can be calculated using \citet{shapiro1986asymptotic}.

\begin{proposition}\label{prop: efficiency}
	Under model \eqref{eq: mixed_model} and assume envelope conditions (i)$^{*}$ and (ii)$^{*}$ hold, 
	then $
	\sqrt{n}(\widehat{{\bm \theta}}_{mix\cdot em}-{\bm \theta})\xrightarrow{d} N(\bm 0, \V),
	$ and $\sqrt{n}(\widehat{{\bm \theta}}_{mix\cdot env}-{\bm \theta})\xrightarrow{d} N(\bm 0, \V_{mix\cdot env})$ where $\V_{mix\cdot env} = \mathbf G(\mathbf G^T\V\mathbf G)^\dagger\mathbf G$, the form of $\mathbf V$ is given in the appendix, and $\mathbf G$ is given by
	$$\begin{pmatrix}
	\mathbf I_p\otimes\bm\Gamma & \bm\eta^T\otimes\mathbf I_r & \bm 0 & \bm 0 & \bm 0\\
	\bm0  & 2\mathbf C_r(\bm\Gamma\bm\Omega\otimes\mathbf I_r - \bm\Gamma\otimes\bm\Gamma_0\bm\Omega_0\bm\Gamma_0^T) & \mathbf C_r(\bm\Gamma\otimes\bm\Gamma)\mathbf E_u & \mathbf C_r(\bm\Gamma_0\otimes\bm\Gamma_0)\mathbf E_{r-u} & \bm 0\\
	\bm 0 & \bm 0 & \bm 0  & \bm 0& \mathbf I_{qr(qr+1)/2}
	\end{pmatrix},$$
	Moreover, $\mathbf V^{-\frac{1}{2}}(\mathbf V - \mathbf V_0)\mathbf V^{-\frac{1}{2}} =  \mathbf I - \mathbf V^{-\frac{1}{2}} \mathbf G(\mathbf G^T \mathbf V^{-1}\mathbf G)^{\dagger}\mathbf G^T \mathbf V^{-\frac{1}{2}}\geq 0,$ so the mixed effects envelope always has no larger asymptotic variance. 
\end{proposition}

In order to provide some insights on occasions where our estimator can be efficient as compared with the standard method, we compare $\text{avar}[\sqrt n\text{vec}(\widehat{\bm \beta})]$ using the mixed envelope model with the standard model under a relatively simple setting. Specifically, we set $r = 2$, $J = 2$, $p = 1$, $\mathbf Z_{i,j}=1$ for all $i,j$,  $\bm\Sigma_{\bm\varepsilon} = \begin{pmatrix}
	\sigma_1^2 & 0\\0 & \sigma_0^2
	\end{pmatrix}$, $\bm\Sigma_{\mathbf b} = \begin{pmatrix}
	\sigma_b^2 & 0\\
	0 & \sigma_b^2
	\end{pmatrix}$, $\bm\Gamma = (1,0)^T$, $\bm\Gamma_0 = (0, 1)^T$ and $\eta = 1$.
	In this specific case, we have the close form formula $$\text{avar}[\sqrt{n} \text{vec}(\widehat{\bm\beta}_{mix\cdot em})] = \begin{pmatrix}
	\dfrac{\sigma_1^2(\sigma_1^2 + 2\sigma_b^2)}{\sigma_b^2\sigma_{x_1}^2 + \sigma_1^2\sigma_{x_2}^2} & 0\\
	0 & \dfrac{\sigma_0^2(\sigma_0^2 + 2\sigma_b^2)}{\sigma_b^2\sigma_{x_1}^2 + \sigma_0^2\sigma_{x_2}^2}
	\end{pmatrix},$$ 
	and $$\text{avar}[\sqrt{n} \text{vec}(\widehat{\bm\beta}_{mix\cdot env})]= \begin{pmatrix}
	\dfrac{\sigma_1^2(\sigma_1^2 + 2\sigma_b^2)}{\sigma_b^2\sigma_{x_1}^2 + \sigma_1^2\sigma_{x_2}^2} & 0\\
	0 & \sigma_{\beta_2}^2
	\end{pmatrix},$$
	where $\sigma_{\beta_2}^2 = \left[\dfrac{\sigma_b^2\sigma_{x_1}^2 + \sigma_0^2\sigma_{x_2}^2}{\sigma_0^2(\sigma_0^2 + 2\sigma_b^2)} + \dfrac{4(\sigma_1^2 - \sigma_0^2)^2(\sigma_1^2\sigma_0^2 + 2\sigma_1^2\sigma_b^2 + 2\sigma_0^2\sigma_b^2 + 2\sigma_b^4)}{\sigma_1^2\sigma_0^2(\sigma_1^2 + 2\sigma_b^2)(\sigma_0^2+2\sigma_b^2)}\right]^{-1}$,  $\sigma_{x_1}^2 = \sum_{i=1}^{n}(x_{i1} - x_{i2})^2/n$ and $\sigma_{x_2}^2 = \sum_{i=1}^{n}(x_{i1}^2 + x_{i2}^2)/n$. As long as  $\sigma^2_1 \neq \sigma_0^2$ and $\sigma_0^2>0$,
	$$\sigma_{\beta_2}^2< \dfrac{\sigma_0^2(\sigma_0^2 + 2\sigma_b^2)}{\sigma_b^2\sigma_{x_1}^2 + \sigma_0^2\sigma_{x_2}^2}.$$
	The ratio $$\dfrac{\text{avar}[\sqrt{n} \text{vec}(\widehat{\bm\beta}_{mix\cdot em}^{(2)})]}{\text{avar}[\sqrt{n} \text{vec}(\widehat{\bm\beta}_{mix\cdot env}^{(2)})]} = 1 + \dfrac{4(\sigma_1^2 - \sigma_0^2)^2(\sigma_1^2\sigma_0^2 + 2\sigma_1^2\sigma_b^2 + \sigma_0^2\sigma_b^2 + 2\sigma_b^4)}{\sigma_1^2(\sigma_1^2 + 2\sigma_b^2)(\sigma_b^2\sigma_{x_1}^2+\sigma_0^2\sigma_{x_2}^2)}$$ tends to $+\infty$ as $\sigma_0^2\rightarrow +\infty$.
	Therefore, the efficiency gain is large when $\sigma_0^2$ is large relative to $\sigma_1^2$. Consequently, in this case, fewer samples are needed to detect the same effect size for our method as compared with the standard EM.

	
	The consistency and efficiency gain of the mixed effects envelope estimator in Proposition \ref{prop: efficiency} is derived based on the normality of the error and random effect. In the next proposition, we justify the $\sqrt{n}$-consistency of $\widehat{\bm \theta}_{mix\cdot env}$ without the normality conditions on the  error and random effect. 
	\begin{proposition}\label{prop: robustness}
		If the error $\bm\varepsilon_{ij}$ and random effect $\mathbf b_i$ have finite $(4+\delta)$-th moments for some $\delta>0$, and the regularity conditions in the appendix hold, then 
		$\sqrt{n}(\widehat{{\bm \theta}}_{mix\cdot em}-{\bm \theta})\xrightarrow{d} N(\bm 0, \widetilde{\V})$, and $\sqrt{n}(\widehat{{\bm \theta}}_{mix\cdot env}-{\bm \theta})\xrightarrow{d} N(\bm 0, \widetilde{\V}_{mix\cdot env})$ for some covariance matrices $\widetilde{\V}$ and $\widetilde{\V}_{mix\cdot env}$. In addition, we have $\widetilde{\V}_{mix\cdot env} = \mathbf G(\mathbf G^T\mathbf J\mathbf G)^\dagger\mathbf G^T\mathbf J\widetilde{\mathbf V}\mathbf J \mathbf G(\mathbf G^T\mathbf J\mathbf G)^\dagger\mathbf G^T$. The definition of $\mathbf J$ is given in the Appendix.
	\end{proposition}
	
	\section{Simulations}\label{sec: simulation}
	In this section, we carry out simulations to compare the finite sample efficiency of our estimator with the standard EM method, the response envelope method and the response PLS method using the SIMPLS algorithm. The response PLS is a counterpart algorithm of the standard PLS algorithm but to reduce the dimension reduction rather than that for predictors. A detailed algorithm can be found in \citet{cook2018introduction}. The response envelope and the response PLS methods do not take the time dependency among responses into consideration, but still provide consistent estimators. Although envelopes and PLS are asymptotically equivalent as \citet{cook2013envelopes} suggested, their finite sample properties are different.  We first consider a balanced case by generating a population of $n=50$ individuals, and each has $r=10$ responses measured at each of the 5 time points ($J_i= J =5,\ i = 1,\ldots, 50$). We set 
	$p = 6$ and $q = 2$ for the fixed and random effects.
	
	We generate parameters  ${\bm\Gamma}$ and $\bm\beta_0$ of size $r\times u$ and $r\times p$, where $u=1$. The elements of ${\bm\Gamma}$ and $\bm\beta_0$ are from $U(0,1)$ and $U(-10,10)$. Let $\bm\beta  = \Pro_{\bm\Gamma}\bm\beta_0$. We generate a matrix $\mathbf B$ of dimension $qr\times qr$ with each element from $U(-10, 10)$ and let ${\bm\Sigma}_{\mathbf b} = \mathbf B\mathbf B\T$. Let ${\bm\Omega} = 0.01 \I_u$ and ${\bm\Omega}_0 = 100\I_{(r-u)}$ and let ${\bm\Sigma}_{\bm\varepsilon} = {\bm\Gamma}{\bm\Omega}{\bm\Gamma}^T +{\bm\Gamma_0}{\bm\Omega}_0{\bm\Gamma_0}^T$. 
	For each individual, $\X_{i1},\X_{i2},\X_{i3}$ vary with time, and $\X_{i4},\X_{i5},\X_{i6}$ stay fixed for all the time points, where each predictor (time varying or fixed over time) is independently generated from $U(-10, 10)$. Then, generate $\mathbf Z_i = (\mathbf Z_{i1}\T, \mathbf Z_{i2}\T)\T$, where $\mathbf Z_{i1} = \mathbf 1_{1\times J}$, and each element of $\mathbf Z_{i2}$ follows $U(-10, 10)$. Generate vector $\bm{\varepsilon}_{ij}\in \mathbb{R}^{r}$, where each column follows $N(\bm{0}, {\bm\Sigma}_{\bm\varepsilon})$. Also generate $\mathbf{b}_i\in \mathbb{R}^{r\times q}$, where $\mathrm{vec}(\mathbf b_{i})$ is from the normal distribution $N(\bm{0}, {\bm\Sigma}_{\mathbf b})$. Set $\mathbf {Y}_{ij} = \bm\beta \mathbf {X}_{ij} + \mathbf{b}_i\mathbf Z_{ij} + \bm{\varepsilon}_{ij}$. We then calculate $\widehat{\bm\beta}_{mix\cdot em}$, $\widehat{\bm\beta}_{mix\cdot env}$ $\widehat{\bm\beta}_{env}$ and $\widehat{\bm\beta}_{pls}$, and repeat the above procedure for 100 times.

	\begin{figure}[!ht]
		\caption{Empirical distribution of $\|\widehat{\bm \beta} - \bm\beta\|_2^2$}
		\centering
		\includegraphics[width=.7\linewidth]{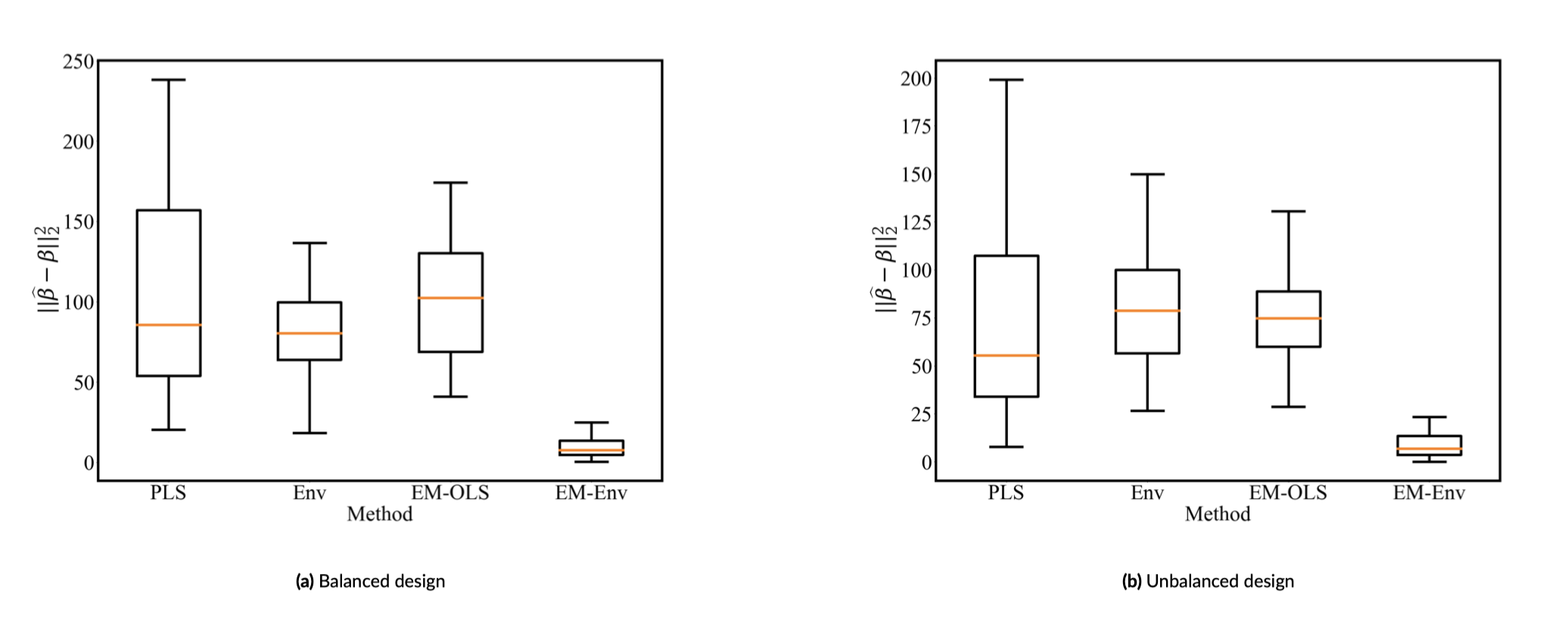}
		\label{simulation}
	\end{figure}

	We compute the square of $l_2$ norm $\|\widehat{\bm \beta} - \bm\beta\|_2^2$ for each simulation. The boxplot of $l_2^2$ error across 100 simulations is given in Figure \ref{simulation}a, where we suppress the outliers to make the figure clean. The mixed envelope estimates are significantly better in terms of both bias and variance than the standard EM estimates. For example, the mean $l_2^2$ error of the mixed envelope estimate for $\bm{\beta}$ is $12.12$, while that is $104.77$, $79.40$ and $105.99$ for the standard EM, response envelope and response PLS estimates respectively. Also, 99 out of 100 of our method selected the correct envelope dimension $u = 1$. 

	Then, we examine the results where $J_i$ is uniformly generated from $\{5, 6, 7, 8, 9\}$. 
	Other steps in the previous simulation remain unchanged. The mean $l_2^2$ error of $\widehat{\bm{\beta}}_{mix\cdot env}$ is $9.30$, while that is $69.11$, $79.05$ and $72.16$ for $\widehat{\bm{\beta}}_{mix\cdot em}$, $\widehat{\bm\beta}_{env}$ and $\widehat{\bm\beta}_{pls}$. Also, the envelope dimension is always correctly estimated as  $\widehat u = 1$. The empirical distribution of $l_2^2$   error is shown in Figure \ref{simulation}b. Our proposed mixed effects envelope estimator has much smaller MSE than the standard EM estimator even in relatively small samples.

	\section{Data Analysis}\label{sec: data_analysis}
	In this section, we apply our proposed method to the Action to Control Cardiovascular Risk in Diabetes (ACCORD) study. The ACCORD randomized-control trial aimed at determining whether cardiovascular disease (CVD) event rates can be reduced in people with diabetes. Participants are between the ages of 40 and 82. All participants have Type 2 diabetes and an especially high risk for heart attack and stroke. 
	
	We are interested in the treatment effect on the quality of life and changes in health outcomes. The responses were collected at four time points, which are 12, 24, 36, 48 months after the beginning of the trial. We consider 2054 participants who responded to the survey, among whom 1156 individuals responded to all surveys. In our analysis, the response variables were treatment satisfaction, depression scale, aggregate physical activity score, aggregate mental score, aggregate interference score, symptom and distress score, systolic blood pressure (SBP), diastolic blood pressure (DBP), and heart rate. The predictors were age and treatment, where we consider the intensive glycemia treatment $(T = 1)$ and the standard glycemia treatment $(T = 0)$.

	\begin{figure}[!ht]
		
		\caption{The empirical cumulative distribution of the ratio between the estimated standard errors of the standard EM and that of our method for ACCORD data.}
		\centering
		\includegraphics[width=.65\linewidth]{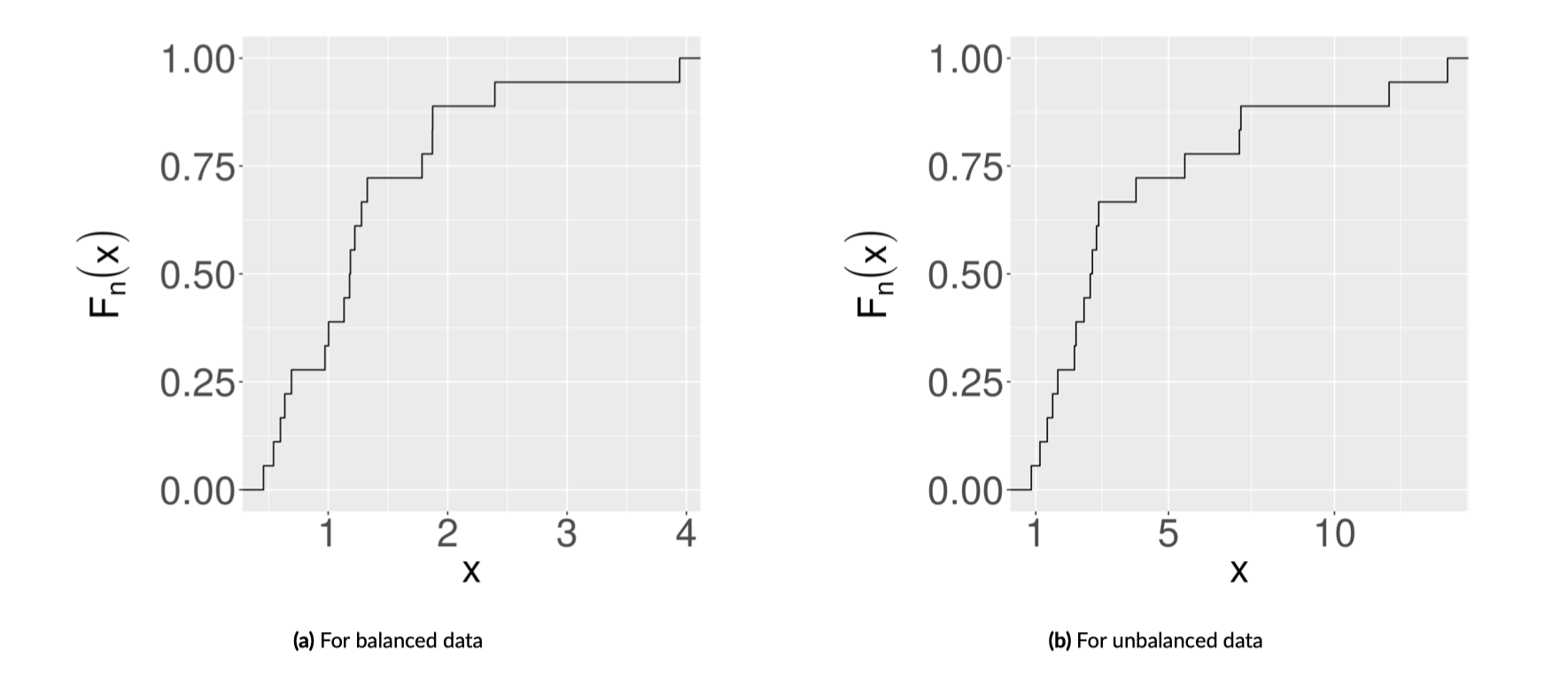}
		\label{ratio1}
	\end{figure}
	
	We first assessed the difference in the quality of life versus glycemia level and age for people who attended all four surveys ($J_i = 4$). All responses except systolic blood pressure and diastolic blood pressure had a missing rate less than 1.5\%. Since the missing rate is low, we imputed the missing data using its mean value. We assume there is only random intercept in our model. The mixed effects envelope method reduced the dimension of the response variable from $r = 9$ to $\widehat u = 4$. The point estimates, bootstrap standard errors, and $p$-values for the regression parameter is given in Table 1 in the Supplementary Material. The magnitude of the point estimates of our method is in general slightly smaller than those of the standard EM. For example, the coefficient for treatment satisfaction with respect to treatment is 0.46 using our method and 0.69 using the standard EM. As the envelope estimate is obtained by projecting the standard estimates onto the envelope direction, the reduction in the magnitude can be interpreted as the noise subtracted from the original estimates. As mentioned in Section \ref{MLE}, the closed form of the standard errors of our method are difficult to obtain. Therefore, we used the nonparametric bootstrap. Figure \ref{ratio1}a  shows the empirical cumulative density distributions of the estimated standard
	errors of the standard EM versus that of our method. The estimated standard errors are in general smaller (on the right hand side of 1 in Figure \ref{ratio1}a) using our method than using the standard EM, which indicates the efficiency gain of our method. The mean ratio of the coefficient standard error using our method over the standard EM is 1.33. That is, to achieve the same mean power among all predictors, our methods require 75.2\% of the original sample.

	Then, we repeated the analysis by including people with less than four surveys, thus $J_i$ varies across each person. In this case, the total number of observations $J_{\bigcdot}$ increases from 1156 to 2054. The missing rate is about the same, so we imputed them using their mean again. The estimated envelope dimension is $\widehat u = 1$. The point estimate, bootstrap standard errors and $p$-values for the regression coefficients are given in Table 2 in the Supplementary Material. It is worth noticing that our method found SBP and DBP corresponding to age significant, whereas the standard method found physical score corresponding to age significant. Figure \ref{ratio1}b shows the empirical distribution of the ratio between the estimated standard errors of the two methods. The mean ratio of the coefficient standard error using our method over the standard EM is 4.07. That is, to achieve the same mean power among all predictors, our methods require 24.6\% of the original sample. All the regression coefficients except symptom and distress score corresponding to age have a smaller standard error, which again shows that our method is more efficient.
	
	\vspace{-5mm}
	\section{Discussion}\label{sec: discussion}
	In this paper, we proposed the mixed effects envelope method to achieve a more efficient estimation than the traditional EM in longitudinal studies. Although this paper is motivated by the repeated measures problem, the mixed effects envelope model can also be used in clustered data to achieve efficiency gain.  For example, patients are nested in physicians, who are in turn nested in clinics. Such clustered data also features in correlations between observations.
	
	The mixed effects model is closely related to the missing data problem since the random effects can be viewed as missing for all observations. Moreover, the missing data techniques may be combined with the mixed effects model to further relax conditions. \citet{ma2019envelope} discussed the envelope method under the ignorable missingness of predictors and covariates. In this paper, we assume that the measures collected at each visit are balanced and  repeated measures may be collected  at different time points across individuals. Such a condition may be violated when a different number of measures are collected every time. One possible solution is to use the union of responses as the balanced response and frame this into a missing data problem. 
	The extension of the mixed effects models with missing data is left as a future research avenue.
	
	In this paper, we considered a heteroscedastic error induced by the mixed effects model. \citet{su2013estimation} proposed an alternative method for another heteroscedastic error covariance structure under the multivariate regression. Both covariance structures allow us to formulate the regression with heteroscedastic variance as a variation of the original envelope model and thus we can use the original computation to obtain the MLE of the likelihood. How to generalize the model with a general heteroscedastic variance structure is left for future research. 
	
	In many contemporary studies and applications, the dimension of data can be much larger than the number of observations. Under such high-dimensional settings, more refined variants of the envelope method are desired. Many studies have adapted the envelope method to high-dimensional settings under sparsity conditions. \citet{su2016sparse} proposed the envelope model for response variable selection under high-dimensional settings. \citet{zhu2020envelope} incorporated the envelope model with partial least squares for high-dimensional regression. We leave the extension of our mixed effects model for high-dimensional data as a future research topic.

	\section*{Acknowledgment}
This research was supported by NSF DMS 1916013, NIH U24-DK-060990 and NIH R01HL155417-01. We would like to thank the Editor, Associate Editor, and two referees for their thorough reviews of the manuscript.
\section*{Supporting Information}
Additional information for this article, including the proof of Proposition 1, 2, 3, 4, graphical illustration of the response envelope, EM-algorithm for the mixed effects envelope model, derivation of $\MakeLowercase{\bm\mu_{\mathbf b_i,t}}$ and $\MakeLowercase{{\bm\Sigma}_{\mathbf b_i,t}}$, the 1-D algorithm, the mixed effects envelope algorithm, and Tables 1 and 2 for the data analysis are available in the supplementary materials.

\bibliography{ref.bib}

\begin{thebibliography}{}

\bibitem[\protect\astroncite{Bi and Qu}{2015}]{bi2015sufficient}
Bi, X. and Qu, A. (2015).
\newblock Sufficient dimension reduction for longitudinal data.
\newblock {\em Statistica Sinica}, 25:787--807.

\bibitem[\protect\astroncite{Cook}{1998}]{cook1998regression}
Cook, R. (1998).
\newblock {\em Regression Graphics: Ideas for Studying Regressions Through
  Graphics}, volume 318.
\newblock John Wiley \& Sons.

\bibitem[\protect\astroncite{Cook}{2018}]{cook2018introduction}
Cook, R. (2018).
\newblock {\em An introduction to envelopes: dimension reduction for efficient
  estimation in multivariate statistics}, volume 401.
\newblock John Wiley \& Sons.

\bibitem[\protect\astroncite{Cook and Forzani}{2009}]{cook2009likelihood}
Cook, R. and Forzani, L. (2009).
\newblock Likelihood-based sufficient dimension reduction.
\newblock {\em Journal of the American Statistical Association}, 104:197--208.

\bibitem[\protect\astroncite{Cook et~al.}{2015}]{cook2015envelopes}
Cook, R., Forzani, L., and Zhang, X. (2015).
\newblock Envelopes and reduced-rank regression.
\newblock {\em Biometrika}, 102:439--456.

\bibitem[\protect\astroncite{Cook et~al.}{2013}]{cook2013envelopes}
Cook, R., Helland, I., and Su, Z. (2013).
\newblock Envelopes and partial least squares regression.
\newblock {\em Journal of Royal Statistical Society: Series B}, 75:851--877.

\bibitem[\protect\astroncite{Cook et~al.}{2010}]{cook2010envelope}
Cook, R., Li, B., and Chiaromonte, F. (2010).
\newblock Envelope models for parsimonious and efficient multivariate linear
  regression.
\newblock {\em Statistica Sinica}, 20:927--960.

\bibitem[\protect\astroncite{Cook and Ni}{2005}]{cook2005sufficient}
Cook, R. and Ni, L. (2005).
\newblock Sufficient dimension reduction via inverse regression: A minimum
  discrepancy approach.
\newblock {\em Journal of the American Statistical Association}, 100:410--428.

\bibitem[\protect\astroncite{Cook and Su}{2013}]{cook2013scaled}
Cook, R. and Su, Z. (2013).
\newblock Scaled envelopes: scale-invariant and efficient estimation in
  multivariate linear regression.
\newblock {\em Biometrika}, 100:939--954.

\bibitem[\protect\astroncite{Cook and Weisberg}{1991}]{cook1991comment}
Cook, R. and Weisberg, S. (1991).
\newblock Comment.
\newblock {\em Journal of the American Statistical Association}, 86:328--332.

\bibitem[\protect\astroncite{Cook and Zhang}{2015a}]{cook2015foundations}
Cook, R. and Zhang, X. (2015a).
\newblock Foundations for envelope models and methods.
\newblock {\em Journal of the American Statistical Association}, 110:599--611.

\bibitem[\protect\astroncite{Cook and Zhang}{2015b}]{cook2015simultaneous}
Cook, R. and Zhang, X. (2015b).
\newblock Simultaneous envelopes for multivariate linear regression.
\newblock {\em Technometrics}, 57:11--25.

\bibitem[\protect\astroncite{Cook and Zhang}{2016}]{cook2016algorithms}
Cook, R. and Zhang, X. (2016).
\newblock Algorithms for envelope estimation.
\newblock {\em Journal of Computational and Graphical Statistics}, 25:284--300.

\bibitem[\protect\astroncite{Dong and Li}{2010}]{dong2010dimension}
Dong, Y. and Li, B. (2010).
\newblock Dimension reduction for non-elliptically distributed predictors:
  second-order methods.
\newblock {\em Biometrika}, 97:279--294.

\bibitem[\protect\astroncite{Eck and Cook}{2017}]{eck2017weighted}
Eck, D. and Cook, R. (2017).
\newblock Weighted envelope estimation to handle variability in model
  selection.
\newblock {\em Biometrika}, 104:743--749.

\bibitem[\protect\astroncite{Fung et~al.}{2002}]{fung2002dimension}
Fung, W., He, X., Liu, L., and Shi, P. (2002).
\newblock Dimension reduction based on canonical correlation.
\newblock {\em Statistica Sinica}, pages 1093--1113.

\bibitem[\protect\astroncite{Hughes and Haran}{2013}]{hughes2013dimension}
Hughes, J. and Haran, M. (2013).
\newblock Dimension reduction and alleviation of confounding for spatial
  generalized linear mixed models.
\newblock {\em Journal of the Royal Statistical Society: Series B (Statistical
  Methodology)}, 75:139--159.

\bibitem[\protect\astroncite{Jones}{2011}]{jones2011bayesian}
Jones, R. (2011).
\newblock Bayesian information criterion for longitudinal and clustered data.
\newblock {\em Statistics in Medicine}, 30:3050--3056.

\bibitem[\protect\astroncite{Li and Dong}{2009}]{li2009dimension}
Li, B. and Dong, Y. (2009).
\newblock Dimension reduction for nonelliptically distributed predictors.
\newblock {\em The Annals of Statistics}, 37:1272--1298.

\bibitem[\protect\astroncite{Li and Wang}{2007}]{li2007directional}
Li, B. and Wang, S. (2007).
\newblock On directional regression for dimension reduction.
\newblock {\em Journal of the American Statistical Association}, 102:997--1008.

\bibitem[\protect\astroncite{Li et~al.}{2005}]{li2005contour}
Li, B., Zha, H., and Chiaromonte, F. (2005).
\newblock Contour regression: a general approach to dimension reduction.
\newblock {\em The Annals of Statistics}, 33:1580--1616.

\bibitem[\protect\astroncite{Li}{1991}]{li1991sliced}
Li, K. (1991).
\newblock Sliced inverse regression for dimension reduction.
\newblock {\em Journal of the American Statistical Association}, 86:316--327.

\bibitem[\protect\astroncite{Li}{1992}]{li1992principal}
Li, K. (1992).
\newblock On principal hessian directions for data visualization and dimension
  reduction: {A}nother application of {S}tein's lemma.
\newblock {\em Journal of the American Statistical Association}, 87:1025--1039.

\bibitem[\protect\astroncite{Li and Zhang}{2017}]{li2017parsimonious}
Li, L. and Zhang, X. (2017).
\newblock Parsimonious tensor response regression.
\newblock {\em Journal of the American Statistical Association},
  112:1131--1146.

\bibitem[\protect\astroncite{Ma et~al.}{2019}]{ma2019envelope}
Ma, L., Liu, L., and Yang, W. (2019).
\newblock Envelope method of ignorable missing data.
\newblock {\em Submitted}.

\bibitem[\protect\astroncite{Magus and Neudecker}{1984}]{magus1984matrix}
Magus, J. and Neudecker, H. (1984).
\newblock Matrix differential calculus with applications to simple, hadamard,
  and kronecker products.
\newblock Technical report.

\bibitem[\protect\astroncite{Park et~al.}{2017}]{park2017groupwise}
Park, Y., Su, Z., and Zhu, H. (2017).
\newblock Groupwise envelope models for imaging genetic analysis.
\newblock {\em Biometrics}, 73:1243--1253.

\bibitem[\protect\astroncite{Pfeiffer et~al.}{2012}]{pfeiffer2012sufficient}
Pfeiffer, R., Forzani, L., and Bura, E. (2012).
\newblock Sufficient dimension reduction for longitudinally measured
  predictors.
\newblock {\em Statistics in Medicine}, 31:2414--2427.

\bibitem[\protect\astroncite{Shao}{2003}]{shao2003mathematical}
Shao, J. (2003).
\newblock {\em Mathematical Statistics}.
\newblock Springer Science \& Business Media.

\bibitem[\protect\astroncite{Shapiro}{1986}]{shapiro1986asymptotic}
Shapiro, A. (1986).
\newblock Asymptotic theory of overparameterized structural models.
\newblock {\em Journal of the American Statistical Association}, 81:142--149.

\bibitem[\protect\astroncite{Su and Cook}{2011}]{su2011partial}
Su, Z. and Cook, R. (2011).
\newblock Partial envelopes for efficient estimation in multivariate linear
  regression.
\newblock {\em Biometrika}, 98:133--146.

\bibitem[\protect\astroncite{Su and Cook}{2012}]{su2012inner}
Su, Z. and Cook, R. (2012).
\newblock Inner envelopes: efficient estimation in multivariate linear
  regression.
\newblock {\em Biometrika}, 99:687--702.

\bibitem[\protect\astroncite{Su and Cook}{2013}]{su2013estimation}
Su, Z. and Cook, R. (2013).
\newblock Estimation of multivariate means with heteroscedastic errors using
  envelope models.
\newblock {\em Statistica Sinica}, 23:213--230.

\bibitem[\protect\astroncite{Su et~al.}{2016}]{su2016sparse}
Su, Z., Zhu, G., Chen, X., and Yang, Y. (2016).
\newblock Sparse envelope model: efficient estimation and response variable
  selection in multivariate linear regression.
\newblock {\em Biometrika}, 103:579--593.

\bibitem[\protect\astroncite{Wu}{1983}]{wu1983convergence}
Wu, C.~J. (1983).
\newblock On the convergence properties of the {EM} algorithm.
\newblock {\em The Annals of statistics}, 11:95--103.

\bibitem[\protect\astroncite{Zhou and He}{2008}]{zhou2008dimension}
Zhou, J. and He, X. (2008).
\newblock Dimension reduction based on constrained canonical correlation and
  variable filtering.
\newblock {\em The Annals of Statistics}, 36:1649--1668.

\bibitem[\protect\astroncite{Zhou et~al.}{2010}]{zhou2010reduced}
Zhou, L., Huang, J., Martinez, J., Maity, A., Baladandayuthapani, V., and
  Carroll, R. (2010).
\newblock Reduced rank mixed effects models for spatially correlated
  hierarchical functional data.
\newblock {\em Journal of the American Statistical Association}, 105:390--400.

\bibitem[\protect\astroncite{Zhu and Su}{2020}]{zhu2020envelope}
Zhu, G. and Su, Z. (2020).
\newblock Envelope-based sparse partial least squares.
\newblock {\em The Annals of Statistics}, 48:161--182.

\bibitem[\protect\astroncite{Zhu et~al.}{2010a}]{zhu2010sufficient}
Zhu, L., Wang, T., Zhu, L., and Ferr{\'e}, L. (2010a).
\newblock Sufficient dimension reduction through discretization-expectation
  estimation.
\newblock {\em Biometrika}, 97:295--304.

\bibitem[\protect\astroncite{Zhu et~al.}{2010b}]{zhu2010dimension}
Zhu, L., Zhu, L., and Feng, Z. (2010b).
\newblock Dimension reduction in regressions via average partial mean
  estimation.
\newblock {\em Journal of the American Statistical Association},
  105:1455--1466.

\end{thebibliography}
\bibliographystyle{apa}

\section*{Supplementary Material}
The Supplementary Material contains the proof of Proposition 1, 2, 3, 4, graphical illustration of the response envelope, EM-algorithm for the mixed effects envelope model, derivation of $\MakeLowercase{\bm\mu_{\mathbf b_i,t}}$ and $\MakeLowercase{{\bm\Sigma}_{\mathbf b_i,t}}$, the 1-D algorithm, the mixed effects envelope algorithm, and Tables 1 and 2 for the data analysis.

\section*{Proof of Proposition 1}
Under model \eqref{eq: classic_noenv_long_bal},
\begin{equation}
\widetilde\Y_i = \onebf_{J}\otimes\alphabf + (\mathbf 1_{J}\otimes\betabf) \X_i + \widetilde \varepsilonbf_i,
\end{equation}
Let 
\begin{align*}
\widetilde\Gammabf = \frac{1}{\sqrt{J}}\left[
\begin{matrix}
\I_r\\
\I_r\\
\vdots\\
\I_r
\end{matrix}
\right],
\widetilde\Gammabf_0 = \frac{1}{\sqrt{J}}\left[
\begin{matrix}
&\I_r &\I_r &\cdots &\I_r\\
&-\I_r &\zerobf&\cdots&\zerobf\\
&\zerobf&-\I_r&\cdots&\zerobf\\
&\vdots&\vdots&\ddots&\vdots\\
&\zerobf&\zerobf&\cdots&-\I_r
\end{matrix}
\right].
\end{align*}
Notice that, $\widetilde\Gammabf_0^T \widetilde \Y_i =(\bm\varepsilon_{i1}\T - \bm\varepsilon_{i2}\T, \ldots, \bm\varepsilon_{i1}\T - \bm\varepsilon_{iJ}\T)\T$, and $\widetilde\Gammabf^T \widetilde \Y_i = \sqrt{J}(\bm\alpha +\bm\beta\X_i+ \bar\varepsilonbf_{i})$, where $\bar\varepsilonbf_{i}=\mathbf b_i + \sum_{j=1}^{J}\varepsilonbf_{ij}/J$. We have $\widetilde\Gammabf_0^T \widetilde \Y_i \indep \X_i$ and 
\begin{align*}
\mathrm{Cov}(\widetilde\Gammabf_0^T \widetilde \Y_i, \widetilde\Gammabf^T \widetilde \Y_i) = \mathrm{Cov}(J\bar\varepsilonbf_{i}\T, (\bm\varepsilon_{i1}\T - \bm\varepsilon_{i2}\T, \ldots, \bm\varepsilon_{i1}\T - \bm\varepsilon_{iJ}\T)\T) = \bm 0,
\end{align*}
which indicates $\widetilde\Gammabf_0^T \widetilde \Y_i \indep \widetilde\Gammabf^T \widetilde \Y_i|\X_i$. Hence, Conditions 1 and 2 are satisfied with $(\widetilde\Gammabf, \widetilde\Gammabf_0)$.

The envelope must be contained in $\mathrm{span}(\widetilde\Gammabf)$, i.e., $\widetilde u\leq r$. Hence we proved Corollary 1. We can further find a semi-orthogonal matrix $\bm\Phi$ with maximum dimension such that $\bm\Phi_0^T\widetilde{\bm\Gamma}^T\Y_i \indep( \bm\Phi^T\widetilde{\bm\Gamma}^T\Y_i, \X_i)$. Thus, by definition, $\widetilde{\bm\Gamma}\bm\Phi = \mathbf 1_J\otimes\bm\Phi$ is the basis matrix for $\mathcal{E}_{\subtildeSigmabfepsilonbf}(\widetilde{\mathcal{B}})$. 

\section*{Proof of Proposition 2}
Under model (1), it is easy to verify that  $$\text{vec}(\Gammabf_0^T\Y_i)|\X_i,\Z_i\sim N(\text{vec}(\Gammabf_0^T\bm\beta\X_i), \mathbf M_{22}) ,$$
$$\text{vec}(\Gammabf^T\Y_i)|\text{vec}(\Gammabf_0^T\Y_i),\X_i,\Z_i\sim N(\bm\mu^{**}, \bm\Sigma^{**}),$$
where $\mathbf M_{11} = \I_{J_i}\otimes(\Gammabf^T\bm\Sigma_{\bm\varepsilon}\Gammabf)  + (\Z_i^T\otimes\bm\Gamma^T)\bm\Sigma_b(\Z_i\otimes\bm\Gamma),$
$\mathbf M_{12} = \I_{J_i}\otimes(\Gammabf^T\bm\Sigma_{\bm\varepsilon}\Gammabf_0)  + (\Z_i^T\otimes\bm\Gamma^T)\bm\Sigma_b(\Z_i\otimes\bm\Gamma_0),$
$\mathbf M_{22} = \I_{J_i}\otimes(\Gammabf_0^T\bm\Sigma_{\bm\varepsilon}\Gammabf_0)  + (\Z_i^T\otimes\bm\Gamma_0^T)\bm\Sigma_b(\Z_i\otimes\bm\Gamma_0)$, $\bm\mu^{**} = \vecc(\Gammabf^T\bm\beta\X_i) + \mathbf M_{12}\mathbf M_{22}\inv\{\vecc(\Gammabf_0^T\Y_i) - \vecc(\Gammabf_0^T\bm\beta\X_i)\},$ and $\bm\Sigma^{**} = \mathbf M_{11} - \mathbf M_{12}\mathbf M_{22}\inv\mathbf M_{12}^T.$ 

Condition 1$^{\circ}$ holds if and only if the distribution of $\Gammabf_0^T\Y_i|\X_i,\Z_i$ is free of $\X_i$ and $\Z_i$, that is,  $\Gammabf_0\T\betabf=\zerobf$, $\Z_i\otimes\bm\Gamma_0 = \zerobf$. Condition 2$^{\circ}$ holds if and only $\mathbf M_{12} = 0$, that is, $\mathbf \I_{J_i}\otimes(\bm\Gamma\T\bm\Sigma_{\bm\varepsilon}\bm\Gamma_0)+(\Z_i\T\otimes\bm\Gamma\T)\bm\Sigma_{\mathbf b}(\Z_i\otimes\bm\Gamma_0)=\zerobf$.

\section*{Proof of Proposition 3}
Let $\widehat{\bm \theta}_{obs\cdot em}$ denote the maximizer of $\bm\beta$, $\bm \Sigma_{\bm\varepsilon}$ and $\bm\Sigma_{\mathbf b}$ in the observed data likelihood in \eqref{obs_data_likelihood}. Similarly, let $\widehat{\bm \theta}_{obs\cdot env}$ denote the maximizer of $\bm\beta$, $\bm \Sigma_{\bm\varepsilon}$ and $\bm\Sigma_{\mathbf b}$ in \eqref{obs_data_likelihood} under additional conditions (i)$^{*}$ and (ii)$^{*}$. Let $\mathbf V_0 = \mathrm{avar}(\widehat{\bm \theta}_{obs\cdot env})$ and $\mathbf V = \mathrm{avar}(\widehat{\bm \theta}_{obs\cdot em})$ denote the asymptotic covariance matrices of the estimators obtained by directly maximizing \eqref{obs_data_likelihood} instead of using EM algorithm. Also, let $\bm \phi = (\bm \eta, \bm \Gamma, \bm \Omega, \bm \Omega_0, \bm\Sigma_{\mathbf b})$ and ${\bm\theta} = (\bm \beta, \bm \Sigma_{\bm\varepsilon}, \bm\Sigma_{\mathbf b})$ denote the parameter under the envelope model and the standard model. 
Let $\{\bm \phi_t\}$ and $\{{\bm \theta}_t\}$ denote the EM sequences, i.e., the parameters sequences we obtain from each EM iteration, of the envelope model and the standard model. By Corollary 1 of \citet{wu1983convergence}, the two EM sequences$ \{\bm \phi_t\}$ and $\{{\bm \theta_t}\}$ converge to their unique maximizer of $L_{obs}$. Hence, in order to prove $\mathbf V_{mix\cdot env}\leq \mathbf V_{mix\cdot em}$, it suffices to prove $\mathbf V_0 \leq \mathbf V$. We found function $\mathbf h$ such that $\mathbf h(\bm \phi) =  (h_1(\bm \phi), h_2(\bm \phi), h_3(\bm \phi))$. Because of the over-parameterization of $\bm \theta$, the gradient matrix $\mathbf G = \dfrac{\partial \mathbf h(\bm \phi)}{\partial \bm \phi^T}$ is not of full rank. By Proposition 4.1 in \citet{shapiro1986asymptotic}, we have
\begin{align*}
\mathbf V_0 = \mathbf G(\mathbf G^T \mathbf V^{-1}\mathbf G)^{\dagger}\mathbf G^T.
\end{align*}
\begin{align*}
\mathbf V - \mathbf V_0 = \mathbf V - \mathbf G(\mathbf G^T \mathbf V^{-1}\mathbf G)^{\dagger}\mathbf G^T = \mathbf V^{\frac{1}{2}}\left[\mathbf I - \mathbf V^{-\frac{1}{2}} \mathbf G(\mathbf G^T \mathbf V^{-1}\mathbf G)^{\dagger}\mathbf G^T \mathbf V^{-\frac{1}{2}}\right]\mathbf V^{\frac{1}{2}}.
\end{align*}
Since $\mathbf I - \mathbf V^{-\frac{1}{2}} \mathbf G(\mathbf G^T \mathbf V^{-1}\mathbf G)^{\dagger}\mathbf G^T \mathbf V^{-\frac{1}{2}}$ is the projection matrix onto the orthogonal complement of $\mathrm{span}(\mathbf V^{-\frac{1}{2}}\mathbf G)$, it is positive semi-definite. Hence, $\mathbf V_0 \leq \mathbf V$. In order to find out the close form of $\mathbf V_0$. Because $\widehat{\bm \theta}_{obs\cdot em}$ is MLE, we can obtain $\mathbf V_0$ by inverting its Fisher information matrix.

The log-likelihood is 
\begin{equation}\label{likelihood}
\begin{aligned}
l(\bm\theta, \Y; \X, \Z) &= C \mh\sum_{i = 1}^n\big[
\log\det(\mathbf I_{J_i}\otimes\bm\Sigma_{\bm\varepsilon} + \mathbf A_i\bm\Sigma_{\mathbf b}\mathbf A_i\T) - \{\mathrm{vec}(\mathbf Y_i) - \text{vec}(\bm \alpha\otimes \onebf^{T}_{J_i}) - \mathrm{vec}(\bm\beta\X_i)\}\T\\
&\quad(\mathbf I_{J_i}\otimes\bm\Sigma_{\bm\varepsilon} + \mathbf A_i\bm\Sigma_{\mathbf b}\mathbf A_i\T)\inv\{\mathrm{vec}(\mathbf Y_i) - \text{vec}(\bm \alpha\otimes \onebf^{T}_{J_i})  - \mathrm{vec}(\bm\beta\X_i)\}\big].
\end{aligned}
\end{equation}

For calculating Fisher, we need the following result in \citet{magus1984matrix}:
\begin{lemma}\label{lemma: kron}
	Let $\mathbf U\in \mathbb{R}^{m\times p}$, $\mathbf V\in \mathbb{R}^{r\times s}$ and $\mathbf X\in\mathbb{R}^{n\times q}$, then 
	$$\dfrac{\partial\text{vec}(\mathbf U\otimes \mathbf V)}{\partial \text{vec}^T(\X)} = (\mathbf I_p\otimes \mathbf G)\dfrac{\partial \text{vec}(\mathbf U)}{\partial\text{vec}^T(\X)} + (\mathbf H\otimes \mathbf I_r)\dfrac{\partial\text{vec}(\mathbf V)}{\partial\text{vec}^T(\X)},$$ 
	where $\mathbf G = (\mathbf K_{sm}\otimes \mathbf I_r)(\mathbf I_m\otimes\text{vec}\mathbf V)$, $\mathbf H = (\mathbf I_p\otimes \mathbf K_{sm})(\text{vec}(\mathbf U)\otimes \mathbf I_s)$, with $\mathbf K_{sm}$ being the commutation matrix in $\mathbb R^{sm\times sm}$, such that for any $\mathbf A\in\mathbb R^{s\times m}$, $\mathbf K_{sm}\text{vec}(\mathbf A) = \text{vec}(\mathbf A^T) $.
\end{lemma} 

Let $l_i$ denote the log-likelihood for individual $i$ and $\psi_i(\Y_i,\bm\theta) = \dfrac{\partial l_i}{\partial \bm\theta}$. Denote $\bm\Sigma_i = \I_{J_i}\otimes \bm\Sigma_{\bm\varepsilon}+\mathbf A_i\bm\Sigma_{\mathbf b}\mathbf A_i^T\in\mathbb{R}^{J_ir\times J_ir}$, $\mathbf D_i = \mathrm{vec}(\mathbf Y_i) - \bm\alpha\otimes\mathbf {1}_{J_i} - \mathrm{vec}(\bm\beta\X_i)\in\mathbb{R}^{J_ir}$. We have $\mathbf C_r\mathbf E_r = \I_{r(r+1)/2}$. Also, denote $\mathbf M_{i1} = \dfrac{\partial \mathrm{vech}(\bm\Sigma_i)}{\partial \mathrm{vech}^T(\bm\Sigma_{\bm\varepsilon})}$, and $\mathbf M_{i2} = \dfrac{\partial \mathrm{vech}(\bm\Sigma_i)}{\partial \mathrm{vech}^T(\bm\Sigma_{\mathbf b})}$. By using Lemma \ref{lemma: kron}, we can also obtain the closed form for $\mathbf M_{i1}$:
\begin{equation*}
\begin{aligned}
\mathbf M_{i1} &= \dfrac{\partial \mathrm{vech}(\bm\Sigma_i)}{\partial \mathrm{vech}^T(\bm\Sigma_{\bm\varepsilon})} = \mathbf C_{J_ir} \dfrac{\partial \mathrm{vec}(\bm\Sigma_i)}{\partial \mathrm{vec}^T(\bm\Sigma_{\bm\varepsilon})}\mathbf E_r = \mathbf C_{J_ir} \{(\mathbf I_{J_i}\otimes\mathbf K_{r,J_i})(\text{vec}(\mathbf I_{J_i})\otimes\mathbf I_r)\otimes\mathbf I_r\}  \mathbf E_r,
\end{aligned}
\end{equation*}
Also, the matrix $\mathbf M_{i2} = \mathbf C_{J_ir}(\mathbf A_i\otimes\mathbf A_i)\mathbf E_{qr}$.

Using the notation above, we have
$$l_i(\bm\beta, \Y_i; \X_i, \Z_i) = \mh \log\det(\bm\Sigma_i) \mh\mathbf D_i^T\bm\Sigma^{-1}\mathbf D_i.$$

By matrix calculus, we have 
$$\dfrac{\partial l_i}{\partial\mathrm{vec}^T(\bm\beta)} =-\mathbf D_i^T\bm\Sigma_{i}^{-1}\dfrac{\partial\mathbf D_i}{\partial \text{vec}^T(\bm\beta)} = \mathbf D_i^T\bm\Sigma_i^{-1}(\X_i^T\otimes \I_r), $$
$$\dfrac{\partial l_i}{\partial\mathrm{vech}^T(\bm\Sigma_{\bm\varepsilon})} = \mh \mathrm{vec}^T({\bm\Sigma_i^{-1}})\mathbf E_{J_ir}\mathbf M_{i1}+\dfrac{1}{2}(\mathbf D_i^T\otimes \mathbf D_i^T)(\bm\Sigma_i^{-1}\otimes \bm\Sigma_i^{-1})\mathbf E_{J_ir}\mathbf M_{i1},$$
$$\dfrac{\partial l_i}{\partial\mathrm{vech}^T(\bm\Sigma_{\mathbf b})} = \mh \mathrm{vec}^T({\bm\Sigma_i^{-1}})\mathbf E_{J_ir}\mathbf M_{i2}+\dfrac{1}{2}(\mathbf D_i^T\otimes \mathbf D_i^T)(\bm\Sigma_i^{-1}\otimes \bm\Sigma_i^{-1})\mathbf E_{J_ir}\mathbf M_{i2},$$
and 
$$\psi_i(\Y_i,\bm\theta) = (\psi_{i1}^T, \psi_{i2}^T, \psi_{i3}^T)^T = (\dfrac{\partial l_i}{\partial\mathrm{vec}^T(\bm\beta)} , \dfrac{\partial l_i}{\partial\mathrm{vech}^T(\bm\Sigma_{\bm\varepsilon})} , \dfrac{\partial l_i}{\partial\mathrm{vech}^T(\bm\Sigma_{\mathbf b})})^T.$$
Then, we can calculate the expression for $\dfrac{\partial \psi_i}{\partial\bm\theta^T} $:
$$\dfrac{\partial\psi_{i1}}{\partial \mathrm{vec}^T(\bm\beta)} = -(\X_i\otimes \I_r)\bm\Sigma_i^{-1}(\X_i^T\otimes \I_r),$$
\begin{equation*}
\begin{aligned}
\dfrac{\partial\psi_{i1}}{\partial \mathrm{vech}^T(\bm\Sigma_{\bm\varepsilon})} &=\{\mathbf D_i^T\otimes(\X_i\otimes \I_r)\}\dfrac{\partial\text{vec}(\bm\Sigma_{i}^{-1})}{\partial \text{vec}^T(\bm\Sigma_{\bm\varepsilon})}\\
&= -\{\mathbf D_i^T\otimes(\X_i\otimes \I_r)\}(\bm\Sigma_i^{-1}\otimes \bm\Sigma_i^{-1})\mathbf E_{J_ir}\mathbf M_{i1},
\end{aligned}
\end{equation*}
$$\dfrac{\partial\psi_{i1}}{\partial \mathrm{vech}^T(\bm\Sigma_{\mathbf b})} = -\{\mathbf D_i^T\otimes(\X_i\otimes \I_r)\}(\bm\Sigma_i^{-1}\otimes \bm\Sigma_i^{-1})\mathbf E_{J_ir}\mathbf M_{i2},$$
\begin{equation*}
\begin{aligned}
\dfrac{\partial\psi_{i2}}{\partial \mathrm{vec}^T(\bm\beta)} &= \dfrac{\partial}{\partial \text{vec}^T(\bm\beta)}\left\{\dfrac{1}{2}\mathbf M_{i1}^T\mathbf E_{J_ir}^T(\bm\Sigma_i^{-1}\otimes \bm\Sigma_i^{-1})(\mathbf D_i\otimes\mathbf D_i)\right\}\\
&= \mh\mathbf M_{i1}^T\mathbf E_{J_ir}^T(\bm\Sigma_i^{-1}\otimes \bm\Sigma_i^{-1})(\mathbf I_{rJ_i}\otimes \mathbf D_i + \mathbf D_i\otimes \mathbf I_{rJ_i})(\mathbf X_i^T\otimes\mathbf I_r),
\end{aligned}
\end{equation*}
$$\dfrac{\partial\psi_{i3}}{\partial \mathrm{vec}^T(\bm\beta)} = \mh\mathbf M_{i2}^T\mathbf E_{J_ir}^T(\bm\Sigma_i^{-1}\otimes \bm\Sigma_i^{-1})(\mathbf I_{rJ_i}\otimes \mathbf D_i + \mathbf D_i\otimes \mathbf I_{rJ_i})(\mathbf X_i^T\otimes\mathbf I_r).$$
In order to calculate $\dfrac{\partial\psi_{i2}}{\partial \mathrm{vech}^T(\bm\Sigma_{\bm\varepsilon})}$, 
{\small\begin{equation*}
\begin{aligned}
\dfrac{\partial\psi_{i2}}{\partial \mathrm{vech}^T(\bm\Sigma_{\bm\varepsilon})} & = \dfrac{1}{2}\mathbf M_{i1}^T\mathbf E_{J_ir}^T(\bm\Sigma_i^{-1}\otimes \bm\Sigma_i^{-1})\mathbf E_{J_ir}\mathbf M_{i1} + \dfrac{1}{2}\{(\mathbf D_i^T\otimes\mathbf D_i^T)\otimes (\mathbf M_{i1}^T\mathbf E_{J_ir}^T)\}\dfrac{\partial}{\partial \text{vech}^T(\bm\Sigma_{\bm\varepsilon})}\text{vec}(\bm\Sigma_{i}^{-1}\otimes\bm\Sigma_{i}^{-1}).\\
\end{aligned}
\end{equation*}
By Lemma \ref{lemma: kron}, we have
\begin{equation*}
\begin{aligned}
\dfrac{\partial}{\partial \text{vech}^T(\bm\Sigma_{\bm\varepsilon})}\text{vec}(\bm\Sigma_{i}^{-1}\otimes\bm\Sigma_{i}^{-1}) &= [\mathbf I_{rJ_i}\otimes\{(\mathbf K_{rJ_i,rJ_i}\otimes\mathbf I_{rJ_i})(\mathbf I_{rJ_i}\otimes\text{vec}(\bm\Sigma_{i}^{-1}))\} + \{(\mathbf I_{rJ_i}\otimes\mathbf K_{rJ_i,rJ_i})(\text{vec}(\bm\Sigma_{i}^{-1})\\
&\qquad\otimes \mathbf I_{rJ_i})\otimes\mathbf I_{rJ_i}\}]\{-(\bm\Sigma_{i}^{-1}\otimes\bm\Sigma_{i}^{-1})\mathbf E_{rJ_i}\mathbf M_{i1}\}.
\end{aligned}
\end{equation*}
Hence, 
\begin{equation*}
\begin{aligned}
\dfrac{\partial\psi_{i2}}{\partial \mathrm{vech}^T(\bm\Sigma_{\bm\varepsilon})} & = \dfrac{1}{2}\mathbf M_{i1}^T\mathbf E_{J_ir}^T(\bm\Sigma_i^{-1}\otimes \bm\Sigma_i^{-1})\mathbf E_{J_ir}\mathbf M_{i1} - \dfrac{1}{2}\{(\mathbf D_i^T\otimes\mathbf D_i^T)\otimes (\mathbf M_{i1}^T\mathbf E_{J_ir}^T)\}[\mathbf I_{rJ_i}\otimes\{(\mathbf K_{rJ_i,rJ_i}\otimes\mathbf I_{rJ_i})(\mathbf I_{rJ_i}\otimes\\
&\qquad\otimes\text{vec}(\bm\Sigma_{i}^{-1}))\} + \{(\mathbf I_{rJ_i}\otimes\mathbf K_{rJ_i,rJ_i})(\text{vec}(\bm\Sigma_{i}^{-1})\otimes \mathbf I_{rJ_i})\otimes\mathbf I_{rJ_i}\}]\{(\bm\Sigma_{i}^{-1}\otimes\bm\Sigma_{i}^{-1})\mathbf E_{rJ_i}\mathbf M_{i1}\},
\end{aligned}
\end{equation*}
Similarly,
\begin{equation*}
\begin{aligned}
\dfrac{\partial\psi_{i2}}{\partial \mathrm{vech}^T(\bm\Sigma_{\mathbf b})} & = \dfrac{1}{2}\mathbf M_{i1}^T\mathbf E_{J_ir}^T(\bm\Sigma_i^{-1}\otimes \bm\Sigma_i^{-1})\mathbf E_{J_ir}\mathbf M_{i2} - \dfrac{1}{2}\{(\mathbf D_i^T\otimes\mathbf D_i^T)\otimes (\mathbf M_{i1}^T\mathbf E_{J_ir}^T)\}[\mathbf I_{rJ_i}\otimes\{(\mathbf K_{rJ_i,rJ_i}\otimes\mathbf I_{rJ_i})(\mathbf I_{rJ_i}\otimes\\
&\qquad\otimes\text{vec}(\bm\Sigma_{i}^{-1}))\} + \{(\mathbf I_{rJ_i}\otimes\mathbf K_{rJ_i,rJ_i})(\text{vec}(\bm\Sigma_{i}^{-1})\otimes \mathbf I_{rJ_i})\otimes\mathbf I_{rJ_i}\}]\{(\bm\Sigma_{i}^{-1}\otimes\bm\Sigma_{i}^{-1})\mathbf E_{rJ_i}\mathbf M_{i2}\},
\end{aligned}
\end{equation*}
\begin{equation*}
\begin{aligned}
\dfrac{\partial\psi_{i3}}{\partial \mathrm{vech}^T(\bm\Sigma_{\varepsilonbf})}& = \dfrac{1}{2}\mathbf M_{i2}^T\mathbf E_{J_ir}^T(\bm\Sigma_i^{-1}\otimes \bm\Sigma_i^{-1})\mathbf E_{J_ir}\mathbf M_{i1} - \dfrac{1}{2}\{(\mathbf D_i^T\otimes\mathbf D_i^T)\otimes (\mathbf M_{i2}^T\mathbf E_{J_ir}^T)\}[\mathbf I_{rJ_i}\otimes\{(\mathbf K_{rJ_i,rJ_i}\otimes\mathbf I_{rJ_i})(\mathbf I_{rJ_i}\otimes\\
&\qquad\otimes\text{vec}(\bm\Sigma_{i}^{-1}))\} + \{(\mathbf I_{rJ_i}\otimes\mathbf K_{rJ_i,rJ_i})(\text{vec}(\bm\Sigma_{i}^{-1})\otimes \mathbf I_{rJ_i})\otimes\mathbf I_{rJ_i}\}]\{(\bm\Sigma_{i}^{-1}\otimes\bm\Sigma_{i}^{-1})\mathbf E_{rJ_i}\mathbf M_{i1}\},\\
\end{aligned}
\end{equation*}
\begin{equation*}
\begin{aligned}
\dfrac{\partial\psi_{i3}}{\partial \mathrm{vech}^T(\bm\Sigma_{\mathbf b})}& = \dfrac{1}{2}\mathbf M_{i2}^T\mathbf E_{J_ir}^T(\bm\Sigma_i^{-1}\otimes \bm\Sigma_i^{-1})\mathbf E_{J_ir}\mathbf M_{i2} - \dfrac{1}{2}\{(\mathbf D_i^T\otimes\mathbf D_i^T)\otimes (\mathbf M_{i2}^T\mathbf E_{J_ir}^T)\}[\mathbf I_{rJ_i}\otimes\{(\mathbf K_{rJ_i,rJ_i}\otimes\mathbf I_{rJ_i})(\mathbf I_{rJ_i}\otimes\\
&\qquad\otimes\text{vec}(\bm\Sigma_{i}^{-1}))\} + \{(\mathbf I_{rJ_i}\otimes\mathbf K_{rJ_i,rJ_i})(\text{vec}(\bm\Sigma_{i}^{-1})\otimes \mathbf I_{rJ_i})\otimes\mathbf I_{rJ_i}\}]\{(\bm\Sigma_{i}^{-1}\otimes\bm\Sigma_{i}^{-1})\mathbf E_{rJ_i}\mathbf M_{i2}\}.\\
\end{aligned}
\end{equation*}}

Hence, $$\dfrac{\partial \psi_i}{\partial\bm\theta^T}  = \begin{pmatrix}
\dfrac{\partial\psi_{i1}}{\partial \mathrm{vec}^T(\bm\beta)} & \dfrac{\partial\psi_{i1}}{\partial \mathrm{vech}^T(\bm\Sigma_{\bm\varepsilon})} & \dfrac{\partial\psi_{i1}}{\partial \mathrm{vech}^T(\bm\Sigma_{\mathbf b})}\\
\dfrac{\partial\psi_{i2}}{\partial \mathrm{vec}^T(\bm\beta)} & \dfrac{\partial\psi_{i2}}{\partial \mathrm{vech}^T(\bm\Sigma_{\bm\varepsilon})} & \dfrac{\partial\psi_{i2}}{\partial \mathrm{vech}^T(\bm\Sigma_{\mathbf b})}\\
\dfrac{\partial\psi_{i3}}{\partial \mathrm{vec}^T(\bm\beta)} & \dfrac{\partial\psi_{i3}}{\partial \mathrm{vech}^T(\bm\Sigma_{\bm\varepsilon})} & \dfrac{\partial\psi_{i3}}{\partial \mathrm{vech}^T(\bm\Sigma_{\mathbf b})}
\end{pmatrix}.$$ 
We can obtain the Fisher information of $\bm\theta$ by taking expectation of $\dfrac{\partial \psi_i}{\partial\bm\theta^T}$ with respective to $\mathbf Y_i$
$$I_i(\bm\theta) = \begin{pmatrix}
I_{11}^i & I_{12}^i & I_{13}^i\\
I_{21}^i & I_{22}^i & I_{23}^i\\
I_{31}^i & I_{32}^i & I_{33}^i\\
\end{pmatrix},$$ where
$$I^{i}_{11} = -\mathbb E\{ (-\X_i\otimes \I_r)^T\bm\Sigma_i^{-1}(\X_i\otimes \I_r) = (\X_i\otimes \I_r)\bm\Sigma_i^{-1}(\X_i^T\otimes \I_r)\},$$
$$I^i_{12} = -\mathbb E \{-\{\mathbf D_i^T\otimes(\X_i\otimes \I_r)\}(\bm\Sigma_i^{-1}\otimes \bm\Sigma_i^{-1})\mathbf E_{J_ir}\mathbf M_{i1}\} = 0,$$
$$I^i_{13} = -\mathbb E \{-\{\mathbf D_i^T\otimes(\X_i\otimes \I_r)\}(\bm\Sigma_i^{-1}\otimes \bm\Sigma_i^{-1})\mathbf E_{J_ir}\mathbf M_{i2}\} = 0,$$
By symmetry of $I_i(\bm\theta)$, $I_{21}^i$ = $I_{31}^i = \bm 0$.
\begin{equation*}
\begin{aligned}
I_{22}^i  &=  -\dfrac{1}{2}\mathbf M_{i1}^T\mathbf E_{J_ir}^T(\bm\Sigma_i^{-1}\otimes \bm\Sigma_i^{-1})\mathbf E_{J_ir}\mathbf M_{i1} + \dfrac{1}{2}\{\text{vec}^T(\bm\Sigma_{i})\otimes (\mathbf M_{i1}^T\mathbf E_{J_ir}^T)\}[\mathbf I_{rJ_i}\otimes\{(\mathbf K_{rJ_i,rJ_i}\otimes\mathbf I_{rJ_i})(\mathbf I_{rJ_i}\otimes\\
&\qquad\otimes\text{vec}(\bm\Sigma_{i}^{-1}))\} + \{(\mathbf I_{rJ_i}\otimes\mathbf K_{rJ_i,rJ_i})(\text{vec}(\bm\Sigma_{i}^{-1})\otimes \mathbf I_{rJ_i})\otimes\mathbf I_{rJ_i}\}]\{(\bm\Sigma_{i}^{-1}\otimes\bm\Sigma_{i}^{-1})\mathbf E_{rJ_i}\mathbf M_{i1}\},
\end{aligned}
\end{equation*}
\begin{equation*}
\begin{aligned}
I_{23}^i  &=  -\dfrac{1}{2}\mathbf M_{i1}^T\mathbf E_{J_ir}^T(\bm\Sigma_i^{-1}\otimes \bm\Sigma_i^{-1})\mathbf E_{J_ir}\mathbf M_{i2} + \dfrac{1}{2}\{\text{vec}^T(\bm\Sigma_{i})\otimes (\mathbf M_{i1}^T\mathbf E_{J_ir}^T)\}[\mathbf I_{rJ_i}\otimes\{(\mathbf K_{rJ_i,rJ_i}\otimes\mathbf I_{rJ_i})(\mathbf I_{rJ_i}\otimes\\
&\qquad\otimes\text{vec}(\bm\Sigma_{i}^{-1}))\} + \{(\mathbf I_{rJ_i}\otimes\mathbf K_{rJ_i,rJ_i})(\text{vec}(\bm\Sigma_{i}^{-1})\otimes \mathbf I_{rJ_i})\otimes\mathbf I_{rJ_i}\}]\{(\bm\Sigma_{i}^{-1}\otimes\bm\Sigma_{i}^{-1})\mathbf E_{rJ_i}\mathbf M_{i2}\},
\end{aligned}
\end{equation*}
\begin{equation*}
\begin{aligned}
I_{32}^i  &=  -\dfrac{1}{2}\mathbf M_{i2}^T\mathbf E_{J_ir}^T(\bm\Sigma_i^{-1}\otimes \bm\Sigma_i^{-1})\mathbf E_{J_ir}\mathbf M_{i1} + \dfrac{1}{2}\{\text{vec}^T(\bm\Sigma_{i})\otimes (\mathbf M_{i2}^T\mathbf E_{J_ir}^T)\}[\mathbf I_{rJ_i}\otimes\{(\mathbf K_{rJ_i,rJ_i}\otimes\mathbf I_{rJ_i})(\mathbf I_{rJ_i}\otimes\\
&\qquad\otimes\text{vec}(\bm\Sigma_{i}^{-1}))\} + \{(\mathbf I_{rJ_i}\otimes\mathbf K_{rJ_i,rJ_i})(\text{vec}(\bm\Sigma_{i}^{-1})\otimes \mathbf I_{rJ_i})\otimes\mathbf I_{rJ_i}\}]\{(\bm\Sigma_{i}^{-1}\otimes\bm\Sigma_{i}^{-1})\mathbf E_{rJ_i}\mathbf M_{i1}\},
\end{aligned}
\end{equation*}
\begin{equation*}
\begin{aligned}
I_{33}^i  &=  -\dfrac{1}{2}\mathbf M_{i2}^T\mathbf E_{J_ir}^T(\bm\Sigma_i^{-1}\otimes \bm\Sigma_i^{-1})\mathbf E_{J_ir}\mathbf M_{i2} + \dfrac{1}{2}\{\text{vec}^T(\bm\Sigma_{i})\otimes (\mathbf M_{i2}^T\mathbf E_{J_ir}^T)\}[\mathbf I_{rJ_i}\otimes\{(\mathbf K_{rJ_i,rJ_i}\otimes\mathbf I_{rJ_i})(\mathbf I_{rJ_i}\otimes\\
&\qquad\otimes\text{vec}(\bm\Sigma_{i}^{-1}))\} + \{(\mathbf I_{rJ_i}\otimes\mathbf K_{rJ_i,rJ_i})(\text{vec}(\bm\Sigma_{i}^{-1})\otimes \mathbf I_{rJ_i})\otimes\mathbf I_{rJ_i}\}]\{(\bm\Sigma_{i}^{-1}\otimes\bm\Sigma_{i}^{-1})\mathbf E_{rJ_i}\mathbf M_{i2}\},
\end{aligned}
\end{equation*} 
this is because $\mathbb E(\mathbf D_i) = 0$ and $\mathbb E(\mathbf D_i\otimes\mathbf D_i) = \text{vec}(\bm\Sigma_{\mathbf b})$.

The $I_i(\bm\theta)$ we obtained is when assuming $\X_i$ and $\Z_i$ are fixed. 
Since $\widehat{\bm\theta}_{mix\cdot em}$ is the MLE with regularity conditions satisfied, we have
$$\text{Var}(\widehat{\bm\theta}_{mix\cdot em}) = \left\{\sum_{i = 1}^nI_i(\bm\theta)\right\}^{-1}.$$
Let $\bm\phi = (\text{vec}^T(\bm\eta), \text{vec}^T(\bm\Gamma), \text{vech}^T(\bm\Omega), \text{vech}^T(\bm\Omega_0),  \text{vech}^T(\bm\Sigma_{\mathbf b}))^T$, $\mathbf h(\bm\phi) = (\text{vec}^T(\bm\Gamma\bm\eta), \text{vech}^T(\bm\Gamma\bm\Omega\bm\Gamma^T+\bm\Gamma_0\bm\Omega_0\bm\Gamma_0^T), \text{vech}^T(\bm\Sigma_{\mathbf b}))^T$, and $\mathbf G = \dfrac{\partial \mathbf h(\bm\phi)}{\partial \bm\phi^T}$. Then, we have 
$$\mathbf G = \begin{pmatrix}
\mathbf I_p\otimes\bm\Gamma & \bm\eta^T\otimes\mathbf I_r & \bm 0 & \bm 0 & \bm 0\\
\bm0  & 2\mathbf C_r(\bm\Gamma\bm\Omega\otimes\mathbf I_r - \bm\Gamma\otimes\bm\Gamma_0\bm\Omega_0\bm\Gamma_0^T) & \mathbf C_r(\bm\Gamma\otimes\bm\Gamma)\mathbf E_u & \mathbf C_r(\bm\Gamma_0\otimes\bm\Gamma_0)\mathbf E_{r-u} & \bm 0\\
\bm 0 & \bm 0 & \bm 0  & \bm 0& \mathbf I_{qr(qr+1)/2}
\end{pmatrix}.$$
Hence, $\text{Var}(\widehat{\bm\theta}_{mix\cdot env}) = \mathbf V_0 = \mathbf G(\mathbf G^T\mathbf V\mathbf G)^\dagger \mathbf G^T$, where $\mathbf V = \sum_{i = 1}^nI_i(\bm\theta).$

\section*{Proof of Proposition 4}
Since the mixed effects envelope model is overparameterized, we will use Proposition 4.1 of \citet{shapiro1986asymptotic} to prove Proposition 4. We will check their conditions. For convenience, we match Shapiro's notations in our context. Shapiro's $\mathbf x$ in our context is $\widehat{{\bm \theta}}_{mix\cdot em} = (\widehat{{\bm \beta}}_{mix\cdot em}, \widehat{\bm \Sigma}_{\bm\varepsilon\cdot mix\cdot em}, \widehat{\bm \Sigma}_{\mathbf b\cdot mix\cdot em})$. We need to show the $\sqrt n$-consistency and asymptotical normality of $\mathbf x$. We assume the following regularity conditions: the error $\bm\varepsilon_{ij}$ and random effect $\mathbf b_i$ have finite $(4+\delta)$-th moment, $\sup_i\|\X_i\|<\infty$, $\sup_i\|\Z_i\| < \infty$, $\sup_i J_i <\infty$, $\inf_i\det(\mathbf I_{J_i}\otimes\bm\Sigma_{\bm\varepsilon} + \mathbf A_i\bm\Sigma_{\mathbf b}\mathbf A_i\T)>0$, $\liminf_n\lambda_-[n^{-1}\var(s_n(\bm\theta))]>0$ and $\liminf_n\lambda_-[n^{-1}\mathbf M_n(\bm\theta)]>0$, where $\lambda_-[\mathbf A]$ denote the smallest eigenvalue of the matrix $\mathbf A$, $\mathbf M_n(\bm\theta)=-\mathbb{E}\Big\{\dfrac{\partial^2 l}{\partial \bm\theta\partial \bm\theta^T}\Big\}$, and $l$ is the log-likelihood when the error is normally distributed.

The estimator $\widehat{{\bm \theta}}_{mix\cdot em}$ is obtained by maximizing the following misspecified log-likelihood:
\begin{equation*}
	\begin{aligned}
	l(\bm\theta, \Y; \X, \Z) &= C \mh\sum_{i = 1}^n\big[
	\log\det(\mathbf I_{J_i}\otimes\bm\Sigma_{\bm\varepsilon} + \mathbf A_i\bm\Sigma_{\mathbf b}\mathbf A_i\T) - \{\mathrm{vec}(\mathbf Y_i) - \bm\alpha\otimes\mathbf {1}_{J_i} - \mathrm{vec}(\bm\beta\X_i)\}\T\\
	&\quad(\mathbf I_{J_i}\otimes\bm\Sigma_{\bm\varepsilon} + \mathbf A_i\bm\Sigma_{\mathbf b}\mathbf A_i\T)\inv\{\mathrm{vec}(\mathbf Y_i) - \bm\alpha\otimes\mathbf {1}_{J_i} - \mathrm{vec}(\bm\beta\X_i)\}\big].
	\end{aligned}
	\end{equation*}
	Thus $\widehat{{\bm \theta}}_{mix\cdot em}$ is the solution to the generalized estimating equation (GEE)
	$$\dfrac{\partial l}{\partial \bm\theta^T} = \sum_{i=1}^n \dfrac{\partial l_i}{\partial \bm\theta^T} = \sum_{i = 1}^n\psi_i^T(\Y_i,\bm\theta)= 0,$$ 
	where $l_i$ is the misspecified log-likelihood of each observation, and $\psi_i(\Y_i,\bm\theta) = \dfrac{\partial l_i}{\partial \bm\theta}$. Because $\Y_i$ has finite second moment,  $\mathbb{E}\{\psi_i(\Y_i,\bm\theta_0)\} = 0$, where the subscript 0 indicates the true parameter value. We apply Proposition 5.5 and Theorem 5.14 in Shao (2003) to prove consistency and asymptotical normality of $\widehat{{\bm \theta}}_{mix\cdot em}$. 
	
	In order to use Proposition 5.5, we need to show the conditions in Lemma 5.3 in \citet{shao2003mathematical} holds for any compact subset of the parameter space. That is, for any $c>0$ and sequence $\{\mathbf y_i\}_{i=1}^{\infty}$ satisfying $\|\mathbf y_i\|\leq c$, the sequence of functions ${\psi_i(\mathbf y_i,\bm\theta)}$ is equicontinuous on any compact set of the parameter space. It is easy to see that $\dfrac{\partial \psi_i}{\partial\bm\theta^T}$ (derived in the previous subsection) is uniformly bounded in any compact subset $\Theta$ of the parameter space when $\|\mathbf y_i\|\leq c$ if $\sup_i\|\X_i\|<\infty$, $\sup_i\|\Z_i\| < \infty$, $\sup_i J_i <\infty$ and $\sup_i\|\bm\Sigma_i^{-1}\|_\infty<\infty$, where $\|\cdot\|_\infty$ indicates matrix infinity norm. Since $\sup_i\|\bm\Sigma_i^{-1}\|_\infty<\infty$ if and only if $\sup_i\|\Z_i\|<\infty$ and $\inf_i\det(\bm\Sigma_i)>0$, the aforementioned conditions holds under the regularity conditions. Therefore, $\psi_i(\mathbf y_i,\bm\theta)$ is equicontinuous on $\Theta$. Moreover, since $\Y_i$ has finite $(4+\delta)$-th moment, $\mathbb{E}\{\sup_{\bm\theta\in\Theta}\|\psi_i(\Y_i,\bm\theta)\|\}^{2}<\infty$ and $\sup_i\mathbb{E}\|\Y_i\|<\infty$, the conditions in Lemma 5.3 in \citet{shao2003mathematical} holds.
	
	According to Proposition 5.5 of \citet{shao2003mathematical}, we also need to prove $\lim\limits_{n\rightarrow \infty}\dfrac{1}{n}\sum_{i=1}^{\infty}\mathbb{E}\{\psi_i(\mathbf Y_i,\bm\theta)\}=0$ implies $ \bm\theta=\bm\theta_0$. Let  $\bm\theta_{10} = \bm\beta_0, \bm\theta_{20} = \bm\Sigma_{\bm\varepsilon0},\bm\theta_{30} = \bm\Sigma_{\mathbf b0}$ denote the true parameter value, and $\bm\Sigma_{i0} = \I_{J_i}\otimes\bm\theta_{20} + \mathbf A_i\bm\theta_{30}\mathbf A_i^T$. Taking expectation of $\psi_i(\Y_i, \bm\theta)$, we have
	$$\mathbb{E}\{\psi_{i1}(\mathbf Y_i,\bm\theta)\} = \mathrm{vec}^T\{(\bm\theta_{10} - \bm\beta)\mathbf X_i\}{\bm\Sigma}_{i}^{-1}(\X_i\otimes \I_r),$$
	$$\mathbb{E}\{\psi_{i2}(\mathbf Y_i,\bm\theta)\} = \dfrac{1}{2}\mathrm{vec}^T({\bm\Sigma}_{i0}^{-1}\bm\Sigma_i{\bm\Sigma}_{i0}^{-1} - \bm\Sigma_i^{-1})\mathbf E_{J_ir}\mathbf M_{i1},$$
	$$\mathbb{E}\{\psi_{i3}(\mathbf Y_i,\bm\theta)\} = \dfrac{1}{2}\mathrm{vec}^T({\bm\Sigma}_{i0}^{-1}\bm\Sigma_i{\bm\Sigma}_{i0}^{-1} - \bm\Sigma_i^{-1})\mathbf E_{J_ir}\mathbf M_{i2}.$$
	Also,
	$$\mathrm{vec}^T({\bm\Sigma}_{i0}^{-1}\bm\Sigma_i{\bm\Sigma}_{i0}^{-1} - \bm\Sigma_i^{-1})=0 \textrm{ \hspace{2mm}if and only if \hspace{2mm}}\bm\Sigma_i =\bm\Sigma_{i0}. $$
	Because $\X_i$ and $\Z_i$ can be arbitrary,
	$$\bm\Sigma_i = \bm\Sigma_{i0} \textrm{ \hspace{2mm}if and only if \hspace{2mm}}\bm\Sigma_{\bm\varepsilon} =\bm\theta_{20} \textrm{ \hspace{2mm}and \hspace{2mm}}\bm\Sigma_{\mathbf b} =\bm\theta_{30}. $$
	Therefore,
	$$\lim\limits_{n\rightarrow \infty}\dfrac{1}{n}\sum_{i=1}^{\infty}\mathbb{E}\{\psi_{i1}(\mathbf Y_i,\bm\theta)\}=0 \textrm{ \hspace{2mm}implies \hspace{2mm}} \bm\beta= \bm\theta_{10},$$
	$$\lim\limits_{n\rightarrow \infty}\dfrac{1}{n}\sum_{i=1}^{\infty}\mathbb{E}\{\psi_{i2}(\mathbf Y_i,\bm\theta)\}=0 \textrm{ \hspace{2mm}implies \hspace{2mm}} \bm\Sigma_{\bm\varepsilon} = \bm\theta_{20}, \bm\Sigma_{\mathbf b} = \bm\theta_{30}.$$
	Hence $\lim\limits_{n\rightarrow \infty}\dfrac{1}{n}\sum_{i=1}^{\infty}\mathbb{E}\{\psi_i(\mathbf Y_i,\bm\theta)\}=0$ implies $ \bm\theta=\bm\theta_0$. Since $\widehat{{\bm \theta}}_{mix\cdot em}$ is always $O_p(1)$, by Proposition 5.5 in \citep{shao2003mathematical}, $\widehat{{\bm \theta}}_{mix\cdot em} \overset{p}{\rightarrow}\bm\theta_0$.

	Then we prove asymptotic normality of $\widehat{{\bm \theta}}_{mix\cdot em}$ using Theorem 5.14  of \citep{shao2003mathematical}. Since $\Y_i$ has finite $(4+\delta)$-th moment, $\sup_i\|\psi_i(\Y_i,\bm\theta)\|^{2+\frac{\delta}{2}}<\infty$. Then, if conditions $\liminf_n\lambda_-[n^{-1}\var(s_n(\bm\theta))]>0$ and $\liminf_n\lambda_-[n^{-1}\mathbf M_n(\bm\theta)]>0$ holds, we have $$\sqrt{n}(\widehat{{\bm \theta}}_{mix\cdot em} - \bm\theta_0)\overset{d}{\rightarrow}N(0, \widetilde{\mathbf V}),$$
where $\widetilde{\mathbf V} = \dfrac{1}{n}[\mathbf M_n(\bm\theta_0)]^{-1}\var(s_n(\bm\theta_0))[\mathbf M_n(\bm\theta_0)]^{-1}$.

Shapiro's $\bm\xi$ in our context is $\bm\xi = (\bm\beta, \bm\Sigma_{\bm\varepsilon}, \bm\Sigma_{\mathbf b})$. Following the same technique in \citet{su2012inner} and \citet{cook2015envelopes}, we give the minimum discrepancy function as $f_{MDF} = l_{max} - l$, where $l$ is the misspecified log-likelihood function (\ref{likelihood}), and $l_{max}$ is obtained by substituting $\widehat{\bm\theta}_{mix\cdot em}$ for $\bm{\theta}$ in (\ref{likelihood}). Although $f_{MDF}$ is written in terms of $\bm\theta$ and $\widehat{\bm\theta}_{mix\cdot em}$, there must be one-to-one functions $f_1$ from $\bm\theta$ to $\bm\xi$ and $f_2$ from $\widehat{\bm\theta}_{mix\cdot em}$ to $\mathbf x$ so that $\bm\xi = f_1(\bm\theta)$ and $\mathbf x = f_2(\widehat{\bm\theta}_{mix\cdot em})$.  As $f_{MDF}$ is constructed under the normal likelihood, it satisfies the four conditions required by \citet{shapiro1986asymptotic}. Denote $\bm \phi = (\bm \eta, \bm \Gamma, \bm \Omega, \bm \Omega_0, \bm\Sigma_{\mathbf b})$, $\bm\theta = \mathbf h(\bm\phi)$,  $\mathbf G =\partial \mathbf h(\bm \phi)/\partial \bm \phi^T$ and $\mathbf J = \dfrac{1}{2} \cdot \dfrac{\partial^2 f_{MDF}}{\partial\bm\theta\partial\bm\theta^T} $. Notice that $\mathbf J$ equals the Fisher information matrix for $\bm\theta$ when $\bm\varepsilon$ is normal. 

\vspace{2mm}Because $\widehat{{\bm \theta}}_{mix\cdot em}$ is obtained by minimizing $f_{MDF}$, by Proposition 4.1 of \citet{shapiro1986asymptotic}, 
$$\sqrt{n}(\widehat{{\bm \theta}}_{mix\cdot env} - \bm\theta_0)\overset{d}{\rightarrow}N(0, \widetilde{\V}_{mix\cdot env}),$$ where $\widetilde{\V}_{mix\cdot env} = \mathbf G(\mathbf G^T\mathbf J\mathbf G)^\dagger\mathbf G^T\mathbf J\widetilde{\mathbf V}\mathbf J \mathbf G(\mathbf G^T\mathbf J\mathbf G)^\dagger\mathbf G^T$. If we define the inner product as $\langle \mathbf x_1, \mathbf x_2\rangle_{\mathbf J} = \mathbf x_1^T\mathbf J\mathbf x_2$, then the projection onto $\text{span}(\mathbf B)$ relative to $\mathbf J$ has the matrix representation $\mathbf P_{\mathbf B(\mathbf J)} = \mathbf B(\mathbf B^T\mathbf J\mathbf B)^\dagger\mathbf B^T\mathbf J$. Hence, $\widetilde{\V}_{mix\cdot env} = \mathbf P_{\mathbf G(\mathbf J)}\widetilde{\mathbf V}\mathbf P_{\mathbf G(\mathbf J)}\leq\widetilde{\mathbf V}$.

\section*{Graphical illustration of the fixed effects model}
We present a graphical illustration of the fixed effects model. Using the same data as Section \ref{sec: graph}, except that we assume the random effects $\mathbf b_i$ are observed. In this case, we use $\mathbf Y_i - \mathbf b_i$ as response. That is,  (1) holds for response $\Y_i$ and observations are independent for different $i$. After subtracting $\bbf_i$, (3) holds for response $\Y_i-\bbf_i$ and observations are independent for different $i$ and $j$.

\begin{figure}[!h]
	
	\caption{Graphical illustration of the OLS and the classic envelope estimator when the response is $\Y-\bbf$, i.e., with only fixed effect. The two groups are denoted by triangle and cross dots. }
	\centering
	\begin{subfigure}[b]{0.23\textwidth}
		\includegraphics[width=\textwidth]{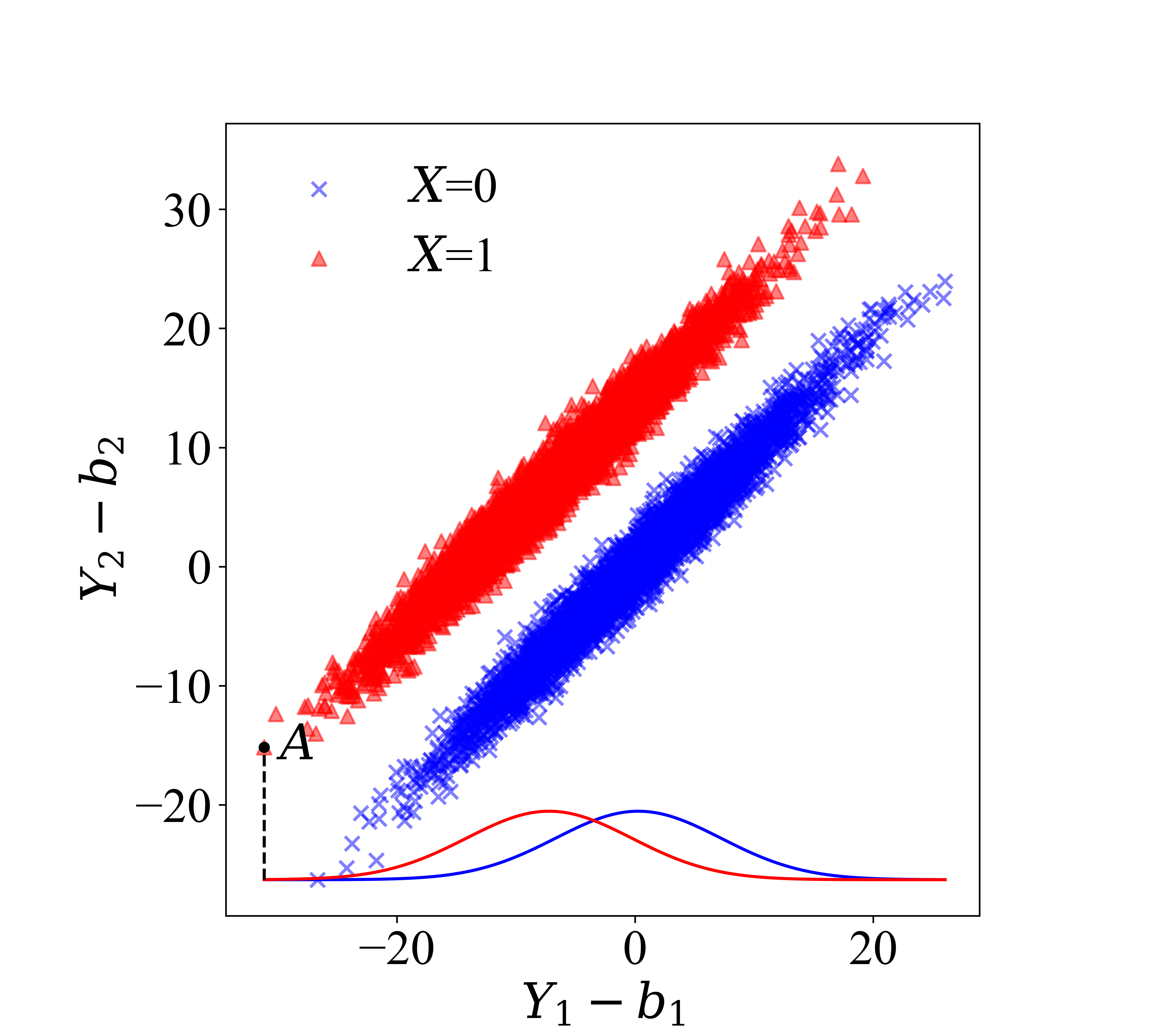}
		\caption{OLS}
		\label{fig1.1}
	\end{subfigure}
	\hspace{2mm}
	\begin{subfigure}[b]{0.23\textwidth}
		\includegraphics[width=\textwidth]{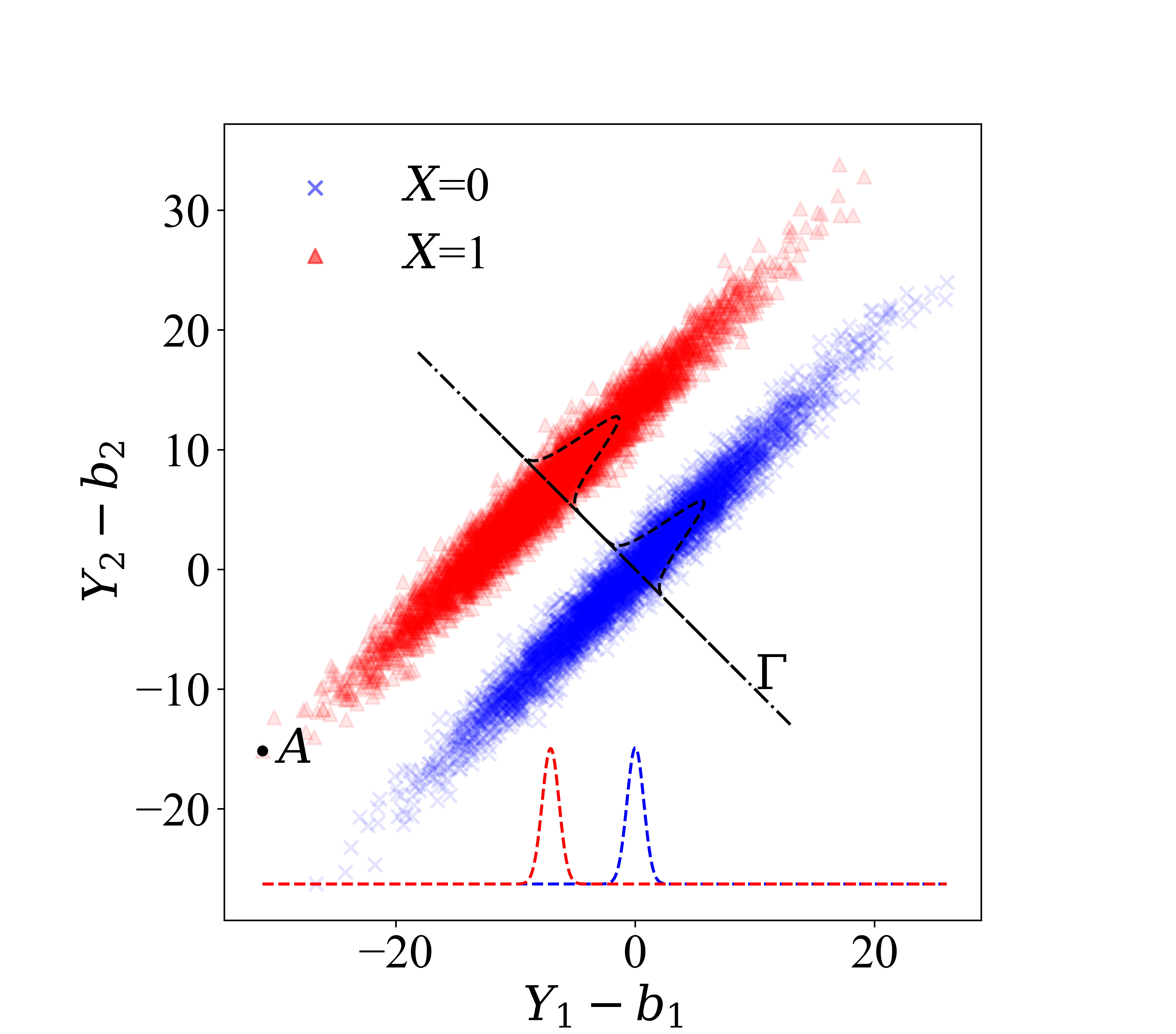}
		\caption{Envelope}
		\label{fig1.2}
	\end{subfigure}
	\label{demon_plot}
\end{figure}

Figure \ref{demon_plot} demonstrates the intuition for the efficiency gain of the classic envelope method in standard multivariate regression with fixed effects only.   In Figure \ref{fig1.1},  the ordinary least square estimator (OLS) is obtained by projecting all the data onto the $Y_1-b_1$ axis, ignoring $Y_2$ completely. The density curves of the two group distributions of $Y_{ij1} - b_{i1}$ are given at the bottom in Figure \ref{fig1.1} and similar curves can be made for $Y_{ij2}- b_{i2}$. The two density curves are not well separated. The OLS of the group difference of $\Y_{ij} - \bbf_{i}$ is $(-7.67, 6.53)^T$ with standard error being $(0.60, 0.60)^T$ and mean square error (MSE) of the group difference being 0.69.

The idea of the envelope method is to reduce noise in the data by projecting each observation onto the direction that can best distinguish the groups. The two groups are well separated along the dashed black line. Also, they  have almost the same distribution in the direction that is orthogonal to the black solid line. Therefore, discarding that part of variation does not sacrifice the information of group difference, but instead, it makes the estimation more efficient. The density curves of the two groups under the envelope estimation are shown at the bottom of Figure \ref{fig1.2} and they have much smaller spreads than the OLS.  The envelope estimate for the group difference of $\Y_{ij} - \bbf_{i}$ is $(-7.10, 7.10)^T$ with standard error $(0.04,0.04)^T$ and  MSE  0.0025.

\section*{EM-algorithm for the mixed effects enevelope model}

Recall the standard EM-algorithm iterates through the following two steps:

(a) E-step. Suppose we have the parameter ${\bm \theta}_t$ from the $t-$th iteration, then we compute the function 
$Q({\bm \theta}|{\bm \theta}_t) = \int l({\bm \theta}|\mathbf X, \mathbf Y, \mathbf Z, \mathbf b)\prod_{i=1}^nf(\mathbf b_i|\mathbf{X}_i,\mathbf{Y}_i,\mathbf Z_i, {\bm \theta}_t)\dif\mathbf b_i,$
where $l$ is the log-likelihood function, and $f$ is the density function. (b) M-step. Find the maximizer of $Q({\bm \theta}|{\bm \theta}_t)$ as ${\bm \theta}_{t+1}$.

To calculate the function $Q({\bm \theta}|{\bm \theta}_t)$ in the E-step, we derive the formulas for both $f(\mathbf b_i|\mathbf X,\mathbf Y,\mathbf Z,{\bm \theta}_t)$ and $l({\bm \theta}|\mathbf X, \mathbf Y, \mathbf Z, \mathbf b)$. Note  
$f({\bbf}_i|\X,\Y,\Z, {\thetabf}_t) = f({\bbf}_i,\Y_i|\X_i, \Z_i, {\thetabf}_t)/f(\Y_i|\X_i, \Z_i, {\thetabf}_t)$, thus, ${\bbf}_i|\X_i, \Y_i, \Z_i, {\thetabf}_t$  also follows a normal distribution $N(\bm\mu_{\bbf_i,t}, {\Sigmabf}_{\bbf_i,t})$, where 
$\bm\mu_{\bbf_i, t} = \big\{\sum_{j = 1}^{J_i}\mathbf A_{ij}\T\bm\Sigma_{\bm\varepsilon, t}\inv(\mathbf Y_{ij} - \bm\alpha_t - \bm\beta_t\mathbf X_{ij})\big\}(\Sigmabf_{\bbf, t}\inv + \sum_{j = 1}^{J_i}\mathbf A_{ij}\T\bm\Sigma_{\bm\varepsilon, t}\mathbf A_{ij})\inv$,
$\bm \Sigma_{\mathbf b_i, t} = (\bm\Sigma_{\mathbf b, t}\inv + \sum_{j = 1}^{J_i}\mathbf A_{ij}\T\bm\Sigma_{\bm\varepsilon, t}\mathbf A_{ij})\inv$. Derivation  of $\bm\mu_{\bbf_i, t}$ and $\bm \Sigma_{\mathbf b_i, t}$ is given  later in the Supplementary Material. 
Therefore,
\begin{equation*}
\begin{split}
&Q(\bm\theta|\bm\theta_t)\\
=& -\frac{rJ_{\bigcdot}}{2}\log(2\pi) - \frac{nqr}{2}\log(2\pi)  - \frac{n}{2}\log|{\bm\Sigma}_{\mathbf b}|- \frac{J_{\bigcdot}}{2}\log|{\bm\Sigma}_{\bm\varepsilon}| \mh\sum_{i=1}^n \E\left\{ \mathrm{vec}(\mathbf b_i)\T{\bm\Sigma}_{\mathbf b}^{-1}\mathrm{vec}(\mathbf b_i)|\X, \Y, \Z; \bm\mu_{\mathbf b_i,t}, {\bm\Sigma}_{\mathbf b_i,t}\right\}\\
&\quad\mh\sum_{i = 1}^n \sum_{j = 1}^{J_i}\E\left\{(\mathbf Y_{ij} -\bm\alpha- \bm\beta \mathbf X_{ij} - {\mathbf b}_i\mathbf Z_{ij})\T{\bm\Sigma}_{\bm\varepsilon}^{-1}(\mathbf Y_{ij} -\bm\alpha- \bm\beta \mathbf X_{ij} - {\mathbf b}_i\mathbf Z_{ij})|\X,\Y,\Z;\bm \mu_{\mathbf b_i,t},{\bm\Sigma}_{\mathbf b_i,t}\right\}\\
=& \sum_{i = 1}^{n}\sum_{j = 1}^{J_i}\mh\left\{(\mathbf Y_{ij} -\bm\alpha- \bm\beta\mathbf X_{ij} - \mathbf A_{ij}\bm\mu_{\mathbf b_i, t})\T \bm\Sigma_{\bm\varepsilon}\inv (\mathbf Y_{ij} -\bm\alpha- \bm\beta\mathbf X_{ij} - \mathbf A_{ij}\bm\mu_{\mathbf b_i, t}) + \tr(\bm\Sigma_{\bm\varepsilon}\inv\mathbf A_{ij}\bm\Sigma_{\mathbf b_i, t}\mathbf A_{ij}^T)\right\}\\
& \quad\mh \sum_{i = 1}^{n}\left\{\bm\mu_{\mathbf b_i, t}\T \bm\Sigma_{\mathbf b}\inv \bm\mu_{\mathbf b_i, t} + \tr(\bm\Sigma_{\mathbf b}\inv\bm\Sigma_{\mathbf b_i, t})\right\} -\dfrac{ J_{\bigcdot}}{2}\log|\bm\Sigma_{\bm\varepsilon}| -\dfrac{n}{2}\log|\bm\Sigma_{\mathbf b}| + C,
\end{split}
\end{equation*}
where $C$ is a constant.

Omitting the constant $C$, $Q({\bm \theta}|{\bm \theta}_t)$ in the above equation can be decomposed as a summation of two parts, i.e., $Q({\bm \theta}|{\bm \theta}_t)=Q_1({\bm\Sigma}_{\mathbf b}|{\bm \theta}_t)+Q_2(\bm\alpha, \bm\beta, {\bm\Sigma}_{\bm\varepsilon}|{\bm \theta}_t)$, where 
\begin{equation*}
\begin{aligned}
Q_1({\bm\Sigma}_{\mathbf b}|{\bm \theta}_t) = &-\frac{n}{2}\log|{\bm\Sigma}_{\mathbf b}| + \sum_{i=1}^n \mh\left\{\bm\mu_{\mathbf b_i,t}\T{\bm\Sigma}_{\mathbf b}^{-1}\bm\mu_{\mathbf b_i,t} + \tr({\bm\Sigma}_{\mathbf b}\inv{\bm\Sigma}_{\mathbf b_i,t})\right\},
\end{aligned}
\end{equation*}
\begin{align*}
Q_2(\bm\alpha, \bm\beta, {\bm\Sigma}_{\bm\varepsilon}|{\bm \theta}_t) =&\sum_{i = 1}^n \sum_{j = 1}^{J_i}\mh \bigg\{(\mathbf Y_{ij} -\bm\alpha- \bm\beta \mathbf X_{ij} - \mathbf A_{ij}\bm \mu_{\mathbf b_i,t})\T{\bm\Sigma}_{\bm\varepsilon}^{-1}(\mathbf Y_{ij} -\bm\alpha- \bm\beta \mathbf X_{ij} - \mathbf A_{ij}\bm\mu_{\mathbf b_i,t})\\
&\quad+\tr({\bm\Sigma}_{\bm\varepsilon}\inv\mathbf A_{ij}{\bm\Sigma}_{\mathbf b_i,t}\mathbf A_{ij}\T)\bigg\} - \frac{J_{\bigcdot}}{2}\log|{\bm\Sigma}_{\bm\varepsilon}|.
\end{align*}
Updates of parameters can be done separately for the two parts. Let $\bm\mu_{\mathbf b,t}=(\bm\mu_{\mathbf b_1,t},\ldots,\bm \mu_{\mathbf b_n,t})$, then $Q_1$ can be written as 
\begin{equation*}
\begin{aligned}
Q_1({\bm\Sigma}_{\mathbf b}|{\bm \theta}_t) 
&= -\frac{n}{2}\log|{\bm\Sigma}_{\mathbf b}| \mh\tr\left\{{\bm\Sigma}_{\mathbf b}\inv(\bm\mu_{\mathbf b,t}\bm\mu_{\mathbf b,t}\T+\sum_{i = 1}^n{\bm\Sigma}_{\mathbf b_i,t})\right\}.
\end{aligned}
\end{equation*}
The update of ${\bm\Sigma}_{\mathbf b}$ is 
${\bm\Sigma}_{\mathbf b,t+1} = \sum_{i=1}^n{\bm\Sigma}_{\mathbf b_i,t}/n+\bm\mu_{\mathbf b,t}\bm\mu_{\mathbf b,t}\T/n.$

Now, we update $\bm\alpha$, $\bm\beta$ and ${\bm\Sigma}_{\bm\varepsilon}$ at the $t+1$ step. 
Under the envelope assumptions, we have ${\bm\Sigma}_{\bm\varepsilon} = {\bm\Sigma}_1 + {\bm\Sigma}_2$, where ${\bm\Sigma}_1={\bm\Gamma}{\bm\Omega}{\bm\Gamma}\T$, ${\bm\Sigma}_2={\bm\Gamma}_0{\bm\Omega}_0{\bm\Gamma}_0\T$ with ${\bm\Sigma}_1 {\bm\Sigma}_2 = \mathbf{0}$, and $\bm\beta$ belongs to the subspace spanned by the column vectors in ${\bm\Sigma}_1$. So we have $\bm\beta\T{\bm\Sigma}_2 = \mathbf{0}$. Moreover, we have ${\bm\Sigma}_{\bm\varepsilon}\inv = {\bm\Sigma}_1^\dagger + {\bm\Sigma}_2^\dagger$,  where the superscript `$\dagger$' denotes 
generalized inverse. 

When $\bm\beta$ and ${\bm\Sigma}_{\bm\varepsilon}$ are fixed, the parameter $\bm\alpha$ maximizing $Q_2(\bm\alpha,\bm\beta,
{\bm\Sigma}_{\bm\varepsilon}|{\bm \theta}_t)$ is  $\widehat{\bm\alpha}= \bar{\mathbf Y} - \bm\beta \bar{\mathbf X} - \bar{\bm\mu}_{t}$,
where $\bar{\mathbf Y} = \sum_{ij} \mathbf Y_{ij}/{J_{\bigcdot}} $, $\bar{\mathbf X} = \sum_{ij} \mathbf X_{ij}/{J_{\bigcdot}} $, and $\bar{\bm\mu}_{t} = \sum_{ij} \mathbf A_{ij}\bm\mu_{\mathbf b_i,t}/{J_{\bigcdot}}$.
According to this relationship, $\bm\alpha_{t+1}$ is a function of $\bm\beta_{t+1}$ and ${\bm\Sigma}_{\bm\varepsilon,t+1}$. Substitute this into the formula of $Q_2$, we obtain
\begin{equation*}
\begin{aligned}
Q_2(\bm\alpha, \bm\beta, {\bm\Sigma}_{\bm\varepsilon}|{\bm \theta}_t) = &- \frac{J_{\bigcdot}}{2}\log|{\bm\Sigma}_{\bm\varepsilon}|\mh\sum_{i = 1}^n \sum_{j = 1}^{J_i} \bigg[\left\{\mathbf Y_{ij} - \bm\beta \mathbf X_{ij} - \mathbf A_{ij}\bm\mu_{\mathbf b_i,t} - (\bar{\mathbf Y} - \bm\beta \bar{\mathbf X} - \bar{\bm\mu}_{t})\right\}\T\\
&\cdot{\bm\Sigma}_{\bm\varepsilon}^{-1}\left\{\mathbf Y_{ij} - \bm\beta \mathbf X_{ij} - \mathbf A_{ij}\bm\mu_{\mathbf b_i,t} - (\bar{\mathbf Y} - \bm\beta \bar{\mathbf X} - \bar{\bm\mu}_{t})\right\}+\tr({\bm\Sigma}_{\bm\varepsilon}\inv\mathbf A_{ij}{\bm\Sigma}_{\mathbf b_i,t}\mathbf A_{ij}\T)\bigg].
\end{aligned}
\end{equation*}
Let $\bm \Psi_{\mathbf b, t} = \sum_{i = 1}^{n}\sum_{j = 1}^{J_i}\mathbf A_{ij}{\bm\Sigma}_{\mathbf b_i,t}\mathbf A_{ij}\T$. Also, let $\mathbf{Y}_c = \mathbf{Y} - \bar{\mathbf Y}\otimes\mathbf{1}_{1\times J_{\bigcdot}}$, 
$\mathbf{X}_c = \mathbf{X} - \bar{\mathbf X}\otimes\mathbf{1}_{1\times J_{\bigcdot}}$, $\bm{\mu}_{c,t} = (\mathbf A_{11}\bm\mu_{\mathbf b_1,t}, \ldots, \mathbf A_{nJ_n}\bm\mu_{\mathbf b_n,t}) - \bar{\bm\mu}_{t}\otimes\mathbf{1}_{1\times J_{\bigcdot}}$.
Then we have
\begin{equation*}
	\begin{aligned}
		&Q_2(\bm\alpha, \bm\beta, {\bm\Sigma}_{\bm\varepsilon}|{\bm \theta}_t) \\
		=& -\frac{1}{2}J_{\bigcdot}\log|{\bm\Sigma}_1 + {\bm\Sigma}_2|- \frac{1}{2}\tr\left\{(\mathbf{Y}_c  - \bm\beta \mathbf{X}_c - \bm{\mu}_{c,t})\T ({\bm\Sigma}_1^\dagger + {\bm\Sigma}_2^\dagger)(\mathbf{Y}_c - \bm\beta \mathbf{X}_c - \bm{\mu}_{c,t})+({\bm\Sigma}_1^\dagger + {\bm\Sigma}_2^\dagger)\bm \Psi_{\mathbf b, t}\right\}\\
		= &-\frac{1}{2}J_{\bigcdot}\log\detz{\bm\Sigma}_1  - \frac{1}{2}\tr\left\{(\mathbf{Y}_c - \bm\beta\mathbf{X}_c- \bm{\mu}_{c,t})\T{\bm\Sigma}_1^\dagger(\mathbf{Y}_c - \bm\beta\mathbf{X}_c- \bm{\mu}_{c,t})\right\} -\frac{1}{2} \tr({\bm\Sigma}_1^\dagger\bm \Psi_{\mathbf b, t})\\
		&-\frac{1}{2}J_{\bigcdot}\log\detz{\bm\Sigma}_2 - \frac{1}{2}\tr\left\{(\mathbf{Y}_c - \bm{\mu}_{c,t})\T {\bm\Sigma}_2^\dagger (\mathbf{Y}_c - \bm{\mu}_{c,t})\right\} - \frac{1}{2}\tr({\bm\Sigma}_2^\dagger \bm \Psi_{\mathbf b, t}).
		\end{aligned}
\end{equation*}

When ${\bm\Sigma}_1$, ${\bm\Sigma}_2$ are fixed, $\widehat\betabf$ that maximizes $Q_2$ is 
\begin{equation}
\label{beta_update1}
\widehat{\bm\beta}_{t+1} = \Pro_{{\bm\Sigma}_1}\mathbf{U}_{c,t}\mathbf{X}_c\T(\mathbf{X}_c \mathbf{X}_c\T)\inv,
\end{equation}
where $\Pro_{{\bm\Sigma}_1}$ denotes the projection matrix on the space spanned by the column vectors of ${\bm\Sigma}_1$, i.e., $ \Pro_{{\bm\Sigma}_1}={{\bm\Sigma}_1}({{\bm\Sigma}_1}\T{{\bm\Sigma}_1})^{-1}{{\bm\Sigma}_1}\T,$ and $\mathbf{U}_{c,t} = \mathbf{Y}_c - \bm{\mu}_{c,t}$. Also, denote $\Q_{{\bm\Sigma}_1}=\I_r- \Pro_{{\bm\Sigma}_1}$, 

Then, we split $Q_2$ into the following two parts:
\begin{equation*}
\begin{aligned}
Q_{2,1}({\bm\Sigma}_1|{\bm \theta}_t) &= -\frac{J_{\bigcdot}}{2}\log\detz{\bm\Sigma}_1 - \frac{1}{2}\tr(\Q_{\mathbf{X}_c}\mathbf{U}_{c,t}\T{\bm\Sigma}_1^\dagger
\mathbf{U}_{c,t}\Q_{\mathbf{X}_c}) -\frac{1}{2} \tr({\bm\Sigma}_1^\dagger\bm \Psi_{\mathbf b, t}),\\
Q_{2,2}({\bm\Sigma}_2|{\bm \theta}_t) &= -\frac{J_{\bigcdot}}{2}\log\detz{\bm\Sigma}_2 - \frac{1}{2}\tr(\mathbf{U}_{c,t}\T{\bm\Sigma}_2^\dagger\mathbf{U}_{c,t})- \frac{1}{2} \tr({\bm\Sigma}_2^\dagger \bm \Psi_{\mathbf b, t}),
\end{aligned}
\end{equation*}
where $\Q_{\mathbf{X}_c} = \mathbf{I} - \Pro_{\mathbf{X}_c}$, and $\mathrm{det}_0(\mathbf A)$ is defined as the product of its non-zero eigenvalues. Suppose ${\bm\Gamma}$ is given, then the maximizers of $Q_{2,1}$ and $Q_{2,2}$ respectively are 
$\widehat{\bm\Sigma}_{1, t+1} = \Pro_{{\bm\Gamma}}(\mathbf{U}_{c,t}\Q_{\mathbf X^{\T}_c}\mathbf{U}_{c,t}\T+\bm \Psi_{\mathbf b, t})\Pro_{{\bm\Gamma}}/J_{\bigcdot},$ $
\widehat{\bm\Sigma}_{2, t+1} =  \Q_{{\bm\Gamma}}(\mathbf{U}_{c,t}\mathbf{U}_{c,t}\T + \bm \Psi_{\mathbf b, t})\Q_{{\bm\Gamma}}/J_{\bigcdot}.
$
The maximized functions are 
$Q_{2,1} = C_1-J_{\bigcdot}\log\detz\left\{\Pro_{{\bm\Gamma}}(\mathbf{U}_{c,t}\Q_{\mathbf X^{\T}_c}\mathbf{U}_{c,t}\T+\bm \Psi_{\mathbf b, t})\Pro_{{\bm\Gamma}}\right\}/2,$
$Q_{2,2} = C_2 - J_{\bigcdot}\log\detz\left\{\Q_{{\bm\Gamma}}(\mathbf{U}_{c,t}\mathbf{U}_{c,t}\T + \bm \Psi_{\mathbf b, t})\Q_{{\bm\Gamma}}\right\}/2.$ 
Hence, $\widehat{\bm \Sigma}_{\varepsilon, t+1} = \widehat{\bm\Sigma}_{1, t+1} + \widehat{\bm\Sigma}_{2, t+1}$. 

The final step is to find the semi-orthogonal matrix ${\bm\Gamma}$ to maximize the function $Q_{2}$, which is equivalent to minimizing the function
\[F(\mathrm{span}({\bm\Gamma})) = \log\det\left\{\Pro_{{\bm\Gamma}}(\mathbf{U}_{c,t}\Q_{\mathbf X^{\T}_c}\mathbf{U}_{c,t}\T + \bm \Psi_{\mathbf b, t})\Pro_{{\bm\Gamma}} + \Q_{{\bm\Gamma}}(\mathbf{U}_{c,t}\mathbf{U}_{c,t}\T+\bm \Psi_{\mathbf b, t})\Q_{{\bm\Gamma}}\right\}.
\]
We only need to identify the span of the column space of ${\bm\Gamma}$ from minimizing the above objective function. We use the 1D algorithm \citep{cook2016algorithms} to obtain $\widehat{\bm \Gamma}$, where $\mathrm{span}(\widehat{\bm \Gamma})$ is a $\sqrt n$-consistent estimator of $\mathrm{span}(\bm\Gamma)$, rather than MLE (more details on 1D algorithm is given later in the Supplementary Material). In our simulation studies in Section \ref{sec: simulation}, our 1D algorithm is feasible and fast converging.

\section*{Derivation of $\MakeLowercase{\bm\mu_{\mathbf b_i,t}}$ and $\MakeLowercase{{\bm\Sigma}_{\mathbf b_i,t}}$}
We derive $\bm\mu_{\mathbf b_i,t}$ and ${\bm\Sigma}_{\mathbf b_i,t}$ in this section. They can be determined from $f({\mathbf b}_i, \mathbf Y_i|\mathbf X_i,\mathbf Z_i, {\bm \theta}_t)$.
\begin{align*}
&f({\mathbf b}_i, \mathbf Y_i|\mathbf X_i,\mathbf Z_i, {\bm \theta}_t) \\
=& 
(2\pi)^{-\frac{qr}{2}}|{\bm\Sigma}_{\mathbf b,t}|^{-\frac{1}{2}}\exp\left\{\mh \mathrm{vec}({\mathbf b}_i)\T{\bm\Sigma}_{\mathbf b,t}^{-1}\mathrm{vec}({\mathbf b}_i)\right\}\\
&\left[\prod_{j = 1}^{J_i} (2\pi)^{-\frac{r}{2}} |{\bm\Sigma}_{\bm\varepsilon,t}|^\mh\exp\left\{\mh(\mathbf Y_{ij} - \bm\alpha_t - \bm\beta_t \mathbf X_{ij} - {\mathbf b}_i\mathbf Z_{ij})\T {\bm\Sigma}_{\bm\varepsilon,t}^{-1}(\mathbf Y_{ij} -\bm\alpha_t- \bm\beta_t \mathbf X_{ij} - {\mathbf b}_i\mathbf Z_{ij})\right\}\right]\\
=&(2\pi)^{-\frac{qr}{2}}|{\bm\Sigma}_{\mathbf b,t}|^{-\frac{1}{2}}\exp\left\{\mh \mathrm{vec}(\mathbf b_i)\T{\bm\Sigma}_{\mathbf b,t}^{-1}\mathrm{vec}(\mathbf b_i)\right\}\\
&\left[\prod_{j = 1}^{J_i} (2\pi)^{-\frac{r}{2}} |{\bm\Sigma}_{\bm\varepsilon,t}|^\mh\exp\left\{\mh(\mathbf Y_{ij} - \bm\alpha_t- \bm\beta_t \mathbf X_{ij} - \mathbf A_{ij}\mathrm{vec}(\mathbf b_i))\T {\bm\Sigma}_{\bm\varepsilon,t}^{-1}(\mathbf Y_{ij} - \bm\alpha_t - \bm\beta_t \mathbf X_{ij} - \mathbf A_{ij}\mathrm{vec}(\mathbf b_i))\right\}\right]\\
\propto& |{\bm\Sigma}_{\mathbf b,t}|^{-\frac{1}{2}}\exp\left\{\mh(\mathrm{vec}(\mathbf b_i) - \bm\mu_{\mathbf b_i, t})\T\bm\Sigma_{\mathbf b_i, t}\inv(\mathrm{vec}(\mathbf b_i) - \bm\mu_{\mathbf b_i, t})\right\}.
\end{align*}
Hence,
\begin{gather*}
\bm\mu_{\mathbf b_i, t} = \left(\bm\Sigma_{\mathbf b, t}\inv + \sum_{j = 1}^{J_i}\mathbf A_{ij}\T\bm\Sigma_{\bm\varepsilon, t}\mathbf A_{ij}\right)\inv\mathbf \sum_{j = 1}^{J_i}\left\{\mathbf A_{ij}\T\bm\Sigma_{\bm\varepsilon, t}\inv(\mathbf Y_{ij} - \bm\alpha_t - \bm\beta_t\mathbf X_{ij})\right\},\\
\bm \Sigma_{\mathbf b_i, t} = \left(\bm\Sigma_{\mathbf b, t}\inv + \sum_{j = 1}^{J_i}\mathbf A_{ij}\T\bm\Sigma_{\bm\varepsilon, t}\mathbf A_{ij}\right)\inv.
\end{gather*}
\section*{The 1D algorithm}
Here we discuss the 1D algorithm to estimate $\mathrm{span}({\bm\Gamma})$. The 1D algorithm was first proposed in \citep{cook2016algorithms} for multivariate regression envelope model. It obtains an estimate of ${\bm\Gamma}$ column-wisely. Specifically, assuming that the envelope dimension $u$ is given, we use the following 1D algorithm to estimate ${\bm\Gamma}$ under the mixed effects model.

\begin{algorithm}[H]\label{1-D}
	\SetAlgoLined
	1. Initialization: $\mathbf g_0 = \mathbf G_0 = 0$\;
	2. For $k = 0, 1, ..., u-1	$,\\
	(a) Let $\mathbf G_k = (\mathbf g_1, ..., \mathbf g_k)$ if $k \geq 1$ and let $(\mathbf G_k, \mathbf G_{0k})$ be an orthogonal basis for $\mathbb{R}^r$.\\
	(b) Define the stepwise objective function\\
	\hspace{1cm}$D_k(\mathbf w) = \log\left(\mathbf w^T\mathbf G_{0k}^{\T}\bm\Sigma_{\varepsilon, t}\mathbf G_{0k} \mathbf w\right) + \log\left\{\mathbf w^T(\mathbf G_{0k}^{\T}\bm\Sigma_{\varepsilon, t}\mathbf G_{0k} + \mathbf G_{0k}^{\T}\bm \beta_t \bm \beta_t^{\T}\mathbf G_{0k})^{-1} \mathbf w\right\}$, \\where $\mathbf w \in \mathbb{R}^{r - k}$.\\
	(c) Solve $\mathbf w_{k+1} = \arg\min_w D_k(\mathbf w) $ subject to a length constraint $\mathbf w^T \mathbf w = 1$.\\
	(d) Define $\mathbf g_{k+1} = \mathbf G_{0k}\mathbf w_{k+1}$ to be the unit length $(k + 1)$th stepwise direction.

	\caption{The 1-D algorithm}
\end{algorithm}
\section*{The mixed effects envelope algorithm}
We combine the 1D algorithm with EM algorithm to obtain an estimator of the mixed effects model under conditions (i)$^{*}$ and (ii)$^{*}$ as follows, where $\delta$ can be chosen depending on the accuracy to achieve.

\begin{algorithm}[H]\label{em_env}
	\SetAlgoLined
	\vspace{2mm}
	\For{k = 1, 2, ..., u}{
		Initialization: $t = 0$, $\bm\Sigma_{\mathbf b, 0} = \I_{qr}$, $\bm\Sigma_{\varepsilon, 0} = \I_r$, $\bm\alpha_0 = \bm 0$, $\bm\beta_0 = \bm 0$, $\bm\theta_0 = (\bm\Sigma_{\mathbf b,0}, \bm\Sigma_{\varepsilon,0}, \bm\alpha_0, \bm\beta_0)$,  $\Delta_0 = \infty$. \\
		\While{$\Delta_t > \delta $}{
			1. Set ${\bm\Sigma}_{\mathbf b_i,t} = \left({\bm\Sigma}_{\mathbf b,t}^{-1} + \sum_{j = 1}^{J_i}\mathbf A_{ij}\T{\bm\Sigma}_{\bm\varepsilon,t}^{-1}\mathbf A_{ij}\right)^{-1}$, where $\mathbf A_{ij} = \I_r\otimes \mathbf Z_{ij}\T$ and $\bm\mu_{\mathbf b_i,t} = \left({\bm\Sigma}_{\mathbf b,t}^{-1}+\sum_{j = 1}^{J_i}\mathbf A_{ij}\T{\bm\Sigma}_{\bm\varepsilon,t}^{-1}\mathbf A_{ij}\right)^{-1}\sum_{j = 1}^{J_i}\left\{\mathbf A_{ij}\T{\bm\Sigma}_{\bm\varepsilon,t}^{-1}(\mathbf Y_{ij} - \bm\alpha_t - \bm\beta_t \mathbf X_{ij})\right\}$. \\
			The update of $\bm \Sigma_{\mathbf b}$ is ${\bm\Sigma}_{\mathbf b,t+1} = \left(\sum_{i=1}^n{\bm\Sigma}_{\mathbf b_i,t}+\bm\mu_{\mathbf b,t}\bm\mu_{\mathbf b,t}\T\right)/N$, where $\bm\mu_{\mathbf b,t} = (\bm\mu_{\mathbf b_1,t}, \ldots, \bm\mu_{\mathbf b_n,t})$.
			
			2. The update of $\bm\alpha$ is $\bm\alpha_{t+1} = \bar{\mathbf Y} - \bm\beta_{t} \bar{\mathbf X} - \bar{\bm\mu}_{t}$,
			where $\bar{\mathbf Y} = \sum_{ij} \mathbf Y_{ij}/{J_{\bigcdot}} $, $\bar{\mathbf X} = \sum_{ij} \mathbf X_{ij}/{J_{\bigcdot}} $, and $\bar{\bm\mu}_{t} = \sum_{ij} \mathbf A_{ij}\bm\mu_{\mathbf b_i,t}/{J_{\bigcdot}}$. 
			
			3. Using the 1D Algorithm to get $\bm\Gamma_{t+1}$. Then, $\bm\beta_{t+1} = \mathbf P_{\bm\Gamma_{t+1}}\mathbf{U}_{c,t}\mathbf{X}_c\T(\mathbf{X}_c \mathbf{X}_c\T)\inv$, where $\mathbf{Y}_c = \mathbf{Y} - \bar{\mathbf Y}\otimes\mathbf{1}_{1\times NJ}$, 
			$\mathbf{X}_c = \mathbf{X} - \bar{\mathbf X}\otimes\mathbf{1}_{1\times NJ}$, $\bm{\mu}_{c,t} = (\mathbf A_{11}\bm\mu_{\mathbf b_1,t}, \ldots, \mathbf A_{nJ_n}\bm\mu_{\mathbf b_n,t}) - \bar{\bm\mu}_{t}\otimes\mathbf{1}_{1\times NJ}$ and $\mathbf U_{c, t} = \mathbf Y_c - \bm\mu_{c, t}$.
			
			4. $\bm\Sigma_{\varepsilon, t+1} = \bm\Sigma_{1, t+1} + \bm\Sigma_{2, t+1}$, where ${\bm\Sigma}_{1, t+1} = \Pro_{{\bm\Gamma_{t+1}}}(\mathbf{U}_{c,t}\Q_{\mathbf X^{\T}_c}\mathbf{U}_{c,t}\T+\bm \Psi_{\mathbf b, t})\Pro_{{\bm\Gamma_{t+1}}} /{J_{\bigcdot}}$, ${\bm\Sigma}_{2, t+1} =  \Q_{{\bm\Gamma_{t+1}}}(\mathbf{U}_{c,t}\mathbf{U}_{c,t}\T +\bm \Psi_{\mathbf b, t})\Q_{{\bm\Gamma_{t+1}}}/{J_{\bigcdot}}$, $\bm \Psi_{\mathbf b, t} = \sum_{i = 1}^{n}\sum_{j = 1}^{J_i}\mathbf A_{ij}{\bm\Sigma}_{\mathbf b_i,t}\mathbf A_{ij}\T$.
			
			5. Set $\Delta_{t+1} = \|\bm \beta_{t + 1} - \bm \beta_t\|_1/\|\bm \beta_{t + 1}\|_1$, $\bm\theta_{t+1} = (\bm \Sigma_{t+1}, \bm \beta_{t + 1}, \bm \rho_{t + 1})$, $t \gets t+1$;
		}
		$\mathrm{BIC}_{k} = -2l({\bm{\widehat \theta}_k};\mathbf Y|\mathbf X) + pu\log n$, where $\bm{\widehat \theta}_k$ and $\bm{\widehat \beta}_k $ are the estimators when the iteration stops.\\
		
	}
	Select $k$ which minimize $\mathrm{BIC}_{k}$, and $\widehat{\bm\beta}_k$ is the mixed effects envelope estimator.
	\vspace{2mm}
	\caption{The mixed effects envelope algorithm}
\end{algorithm}

\begin{table}[!htb]
	\centering
	\caption{The point estimates, bootstrap standard errors and $p-$values for the regression parameter with respect to those patients attended all the measurements}
	\begin{adjustbox}{width=\textwidth,totalheight=\textheight,keepaspectratio}
		\begin{tabular}{c|ccc|ccc}
			\hline
			&\multicolumn{3}{c}{Our Method}&\multicolumn{3}{c}{Standard EM} \\
			\hline
			Corresponding to Treatment& $\bm{\widehat\beta}$ & $\widehat {\mathrm{SE}}$   & $p-$value & $\bm{\widehat\beta}$ & $\widehat {\mathrm{SE}}$  &  $p-$value \\
			\hline
			Treatment Satisfaction & 0.46 & 0.44 & 0.30 & 0.69 & 0.59  &  0.24 \\ 
			Depression Scale & -0.057 & 0.20 & 0.77 & 0.088 & 0.24 &  0.71 \\  
			Physical Score & -2.13$\times 10^{-3}$ & 6.25$\times 10^{-3}$ & 0.73 & -1.81$\times 10^{-3}$ & 0.025  &  0.94 \\ 
			Mental Score & 0.011 & 0.033 & 0.73 & 0.011 & 0.061  &  0.86 \\ 
			Interference Score & 4.57$\times 10^{-4}$ & 1.87$\times 10^{-3}$ & 0.81 & 2.83$\times 10^{-3}$ & 4.48$\times 10^{-3}$  & 0.52 \\ 
			Symptom \& Distress Score & -0.99 & 1.89 & 0.60 & -0.77 & 2.30 & 0.74 \\ 
			SBP & -0.32 & 0.39 & 0.41 & 0.074 & 0.69  & 0.91 \\ 
			DBP & -0.27 & 0.39 & 0.49 & -0.22 & 0.46 & 0.64 \\ 
			Heart Rate & 0.40 & 0.53 & 0.45 & 0.34 & 0.53  &  0.53 \\ 
			\hline	
			Corresponding to Age & $\bm{\widehat\beta}$ & $\widehat {\mathrm{SE}}$   & $p-$value & $\bm{\widehat\beta}$ & $\widehat {\mathrm{SE}}$  &  $p-$value \\
			\hline
			Treatment Satisfaction & 0.24 & 0.10 & 0.02 & 0.23 & 0.046  &  $<0.01$ \\ 
			Depression Scale & -0.039 & 0.019 & 0.04 & -0.071 & 0.013 &  $<0.01$ \\  
			Physical Score & -2.66$\times 10^{-3}$ & 9.95$\times 10^{-4}$ & $<0.01$ & -6.78$\times 10^{-3}$ & 1.87$\times 10^{-3}$  &  $<0.01$ \\ 
			Mental Score & 5.40$\times 10^{-3}$ & 3.65$\times 10^{-3}$ & 0.14 & 0.012 & 4.14$\times 10^{-3}$  &  $<0.01$ \\ 
			Interference Score & 6.93$\times 10^{-5}$ & 2.12$\times 10^{-4}$ & 0.74 & 1.50$\times 10^{-4}$ & 2.71$\times 10^{-4}$  & 0.58 \\ 
			Symptom \& Distress Score & -0.22 & 0.24 & 0.36 & -0.29 & 0.15 & 0.06 \\ 
			SBP & 0.071 & 0.01 & 0.45 & 0.041 & 0.051  & 0.41 \\ 
			DBP & -0.60 & 0.059 & $<0.01$ & -0.61 & 0.035 & $<0.01$\\ 
			Heart Rate & -0.34 & 0.037 & $<0.01$ & -0.34 & 0.036  &  $<0.01$ \\ 
			\hline	
		\end{tabular}
	\end{adjustbox}
	\label{real_data1}
\end{table}

\begin{table}[!thb]
	\centering
	\caption{The point estimates, bootstrap standard errors and $p-$values for the regression parameter with respect to all the patients attended all four measurements}
	\begin{adjustbox}{width=\textwidth,totalheight=\textheight,keepaspectratio}
		\begin{tabular}{c|ccc|ccc}
			\hline
			&\multicolumn{3}{c}{Our Method}&\multicolumn{3}{c}{Standard EM} \\
			\hline
			Corresponding to Treatment& $\bm{\widehat\beta}$ & $\widehat {\mathrm{SE}}$   & $p-$value & $\bm{\widehat\beta}$ & $\widehat {\mathrm{SE}}$  &  $p-$value \\
			\hline
			Treatment Satisfaction & -0.019 & 0.20 & 0.92 & 0.67 & 0.53  &  0.21 \\ 
			Depression Scale & 7.44$\times 10^{-3}$ & 0.080 & 0.93 & 0.13 & 0.18 &  0.48 \\  
			Physical Score & 6.16$\times 10^{-4}$ & 1.41$\times 10^{-3}$ & 0.66 & 3.59$\times 10^{-5}$ & 0.019  &  0.99 \\ 
			Mental Score & -1.36$\times 10^{-3}$ & 0.017 & 0.94 & -0.031 & 0.042  & 0.46 \\ 
			Interference Score & -3.77$\times 10^{-5}$ & 9.39$\times 10^{-4}$ & 0.97 & -6.49$\times 10^{-4}$ & 2.72$\times 10^{-3}$  & 0.81 \\ 
			Symptom \& Distress Score & 0.14 & 1.50 & 0.93 & 1.23 & 1.69 & 0.47 \\ 
			SBP & 8.19$\times 10^{-3}$ & 0.088 & 0.93 & -0.45 & 0.63  & 0.47 \\ 
			DBP & 4.56$\times 10^{-3}$ & 0.050 & 0.93 & -0.60 & 0.36 & 0.10 \\ 
			Heart Rate & 3.06$\times 10^{-3}$ & 0.034 & 0.93 & 0.13 & 0.40  &  0.74 \\ 
			\hline	
			Corresponding to Age & $\bm{\widehat\beta}$ & $\widehat {\mathrm{SE}}$   & $p-$value & $\bm{\widehat\beta}$ & $\widehat {\mathrm{SE}}$  &  $p-$value \\
			\hline
			Treatment Satisfaction & -1.67 & 0.34 & $<0.01$ & 0.23 & 0.51  &  $<0.01$ \\ 
			Depression Scale & -0.65 & 0.12 & $<0.01$ & -0.071 & 0.16 &  $<0.01$ \\  
			Physical Score & -0.061 & 0.011 & 0.40 & -6.78$\times 10^{-3}$ & 0.018  &  $<0.01$ \\ 
			Mental Score & -0.17 & 0.023 & $<0.01$ & 0.012 & 0.050  &  $<0.01$ \\ 
			Interference Score & -0.015 & 1.21$\times 10^{-3}$ & $<0.01$ & 1.50$\times 10^{-4}$ & 3.27$\times 10^{-3}$  & $<0.01$ \\ 
			Symptom \& Distress Score & 1.22 & 2.16 & $<0.01$ & -0.29 & 1.88 & $<0.01$ \\ 
			SBP & 0.72 & 0.18 & $<0.01$ & 0.041 & 0.50  & 0.59 \\ 
			DBP & 0.40 & 0.097 & $<0.01$ & -0.61 & 0.39 & 0.85 \\ 
			Heart Rate & 0.27 & 0.082 & $<0.01$ & -0.34 & 0.45  &  $<0.01$ \\ 
			\hline	
		\end{tabular}
	\end{adjustbox}
	\label{real_data2}
\end{table}

\end{document}